\documentclass{ieeeaccess}
\usepackage{cite}
\usepackage{amsmath,amssymb,amsfonts}
\usepackage{algorithmic}
\usepackage{graphicx}
\usepackage{textcomp}
\usepackage{url}
\usepackage{multirow}
\usepackage{algorithm}
\usepackage{algorithmic}
\usepackage{color}
\usepackage{soul}
\usepackage{xcolor}

\soulregister\cite7
\soulregister\ref7
\soulregister\pageref7

\def\BibTeX{{\rm B\kern-.05em{\sc i\kern-.025em b}\kern-.08em
     T\kern-.1667em\lower.7ex\hbox{E}\kern-.125emX}}

\begin{document}
\history{Date of publication xxxx 00, 0000, date of current version xxxx 00, 0000.}
\doi{10.1109/ACCESS.2017.DOI}

\title{Generic Itemset Mining Based on Reinforcement Learning}
\author{\uppercase{\uppercase{Kazuma Fujioka}\authorrefmark{1} and Kimiaki Shirahama\authorrefmark{2,3}}}
\address[1]{Graduate School of Science and Engineering, Kindai University, 3-4-1, Kowakae, Higashiosaka, Osaka 577-8502, JAPAN (e-mail: kazuma.fujioka@kindai.ac.jp)}
\address[2]{Department of Informatics, Kindai University, 3-4-1, Kowakae, Higashiosaka, Osaka 577-8502, JAPAN (e-mail: shirahama@info.kindai.ac.jp)}
\address[3]{Cyber Informatics Research Institute, Kindai University, 3-4-1, Kowakae, Higashiosaka, Osaka 577-8502, JAPAN}
\tfootnote{This work has been supported in part by Japan Society for the Promotion of Science (JSPS) within Grant-in-Aid for Scientific Research (C) (19K12028).}

\markboth
{K. Fujioka \headeretal: Generic Itemset Mining Based on Reinforcement Learning}
{K. Fujioka \headeretal: Generic Itemset Mining Based on Reinforcement Learning}

\corresp{Corresponding author: Kimiaki Shirahama (e-mail: shirahama@info.kindai.ac.jp).}

\begin{abstract}
One of the biggest problems in itemset mining is the requirement of developing a data structure or algorithm, every time a user wants to extract a different type of itemsets. To overcome this, we propose a method, called \textit{Generic Itemset Mining based on Reinforcement Learning} (GIM-RL), that offers a unified framework to train an agent for extracting any type of itemsets. In GIM-RL, the environment formulates iterative steps of extracting a target type of itemsets from a dataset. At each step, an agent performs an action to add or remove an item to or from the current itemset, and then obtains from the environment a reward that represents how relevant the itemset resulting from the action is to the target type. Through numerous trial-and-error steps where various rewards are obtained by diverse actions, the agent is trained to maximise cumulative rewards so that it acquires the optimal action policy for forming as many itemsets of the target type as possible. In this framework, an agent for extracting any type of itemsets can be trained as long as a reward suitable for the type can be defined. The extensive experiments on mining high utility itemsets, frequent itemsets and association rules show the general effectiveness and one remarkable potential (agent transfer) of GIM-RL. We hope that GIM-RL opens a new research direction towards learning-based itemset mining.
\end{abstract}


\titlepgskip=-15pt

\maketitle

\section{Introduction}
\label{sec:intro}

Much research effort has been made on itemset mining that aims to discover interesting relations among items in a large-scale dataset~\cite{P_Viger_ASO}. The dataset consists of a large number of transactions each of which contains different items. From such a dataset, the goal of itemset mining is to extract interesting sets of items (i.e., itemsets) in terms of a user-specified interestingness measure~\cite{L_Geng}. Depending on the user's needs, different interestingness measures can be chosen to extract, for instance, \textit{Frequent Itemsets} (FIs) consisting of frequently co-occurring items, \textit{Association Rules} (ARs) comprised of correlated items, \textit{High Utility Itemsets} (HUIs) formed by highly profitable items, and so on. From a more general perspective, items in a transaction can be viewed as attributes in one data instance like image, sensor recording and amino acid set. Thus, starting from customer transaction analysis~\cite{R_Agrawal}, itemset mining is utilised in various applications such as image classification, healthcare, bioinformatics and so on~\cite{P_Viger_ASO}.

Itemset mining is difficult because of the huge search space. Assuming that there are $M$ distinct items in a dataset, the search space is defined as the set of all the possible $2^M - 1$ itemsets. The naive approach to examine each of these itemsets is clearly infeasible. Hence, researchers have developed various data structures and algorithms to effectively prune the search space by considering a property specific to a target itemset type. The most popular one is the ``downward closure property'' for FIs, meaning that any subset of an FI must also be frequent~\cite{P_Viger_ASO,R_Agrawal}. Apriori algorithm utilises this property to dramatically prune the search space by ignoring itemsets that contain one or more infrequent subsets of items~\cite{R_Agrawal}. In addition, the downward closure property is used to construct FP-tree (Frequent Patten tree) that offers efficient, hierarchical organisation of items related only to FIs~\cite{J_Han}. Also, considering the lack of the downward closure property for HUIs, upper bound utilities are defined so that the property is maintained among ``potential'' itemsets that have possibilities to be HUIs~\cite{C_Ahmed,V_Tseng}. These upper bound utilities are exploited to construct a tree structure that hierarchically maintains items related only to potential itemsets.

However, the above kind of itemset mining methods are inflexible because a specific data structure or algorithm is needed to extract a different type of itemsets. Especially, one itemset type is defined by an interestingness measure and there exist many such measures as listed in \cite{L_Geng}. It is clearly impractical to develop a data structure or algorithm for each of these itemset types. In other words, diverse types of itemsets remain undiscovered because neither data structure nor algorithm is available. Therefore, one crucial issue in itemset mining is the development of a unified framework to extract different types of itemsets.

For this issue, we propose \textit{Generic Itemset Mining based on Reinforcement Learning} (GIM-RL). The main idea comes from how human searches for a target type of itemsets in a dataset. Most probably, human begins with briefly going through the dataset to get rough knowledge about which items seem to be important (or unimportant) for the target type, which items are related to each other and so on. Then, he/she makes a ``plausible itemset'' consisting of some important items, and examines whether it matches the target type or not. Afterwards, based on knowledge obtained in the past search experiences, human edits the plausible itemset by adding or removing items to create a new plausible itemset. One important point is that human adaptively changes an approach to make plausible itemsets depending on a target type as well as past experiences. Hence, human's itemset search is considered a key for devising a unified itemset mining framework.

We focus on \textit{Reinforcement Learning} (RL) that is rooted in behavioural psychology and provides a framework where an artificial agent interacts with an uncertain environment to adaptively acquire the optimal policy of sequential decision-making~\cite{V_Mnih,V_Mnih2015,V_Lavent}. The agent takes an action at each time step and the environment produces a reward as the response to the action. This trial-and-error step is repeated numerous times to accumulate rewards obtained by various actions at diverse states of the environment. Thereby, RL attempts to find the optimal policy that enables the agent to decide a sequence of actions which maximise cumulative rewards.

RL is utilised in GIM-RL to train an agent for extracting a target type of itemsets from a dataset in the following way: First, the agent performs an action to update an itemset by adding or removing an item. Then, the environment characterised by the dataset generates a reward that expresses how relevant the itemset updated by the agent is to the target type. GIM-RL collects a large number of trial-and-error steps in which the agent sometimes succeeded or sometimes failed in forming itemsets of the target type. By analysing these trial-and-error steps, the agent is trained to have the optimal policy for adding or removing items to form itemsets of the target type. Note that GIM-RL can extract any type of itemsets as long as one can define a reward that appropriately represents the relevance of an itemset to the type. We demonstrate that GIM-RL can be generally applied to mining of HUIs~\cite{V_Tseng,C_Ahmed,M_Liu,W_Song}, FIs~\cite{R_Agrawal,J_Han,P_Viger_ASO} and ARs~\cite{R_Agrawal,P_Viger_ASO,P_Viger_MTA}.

One big by-product of GIM-RL is that it outputs not only extracted itemsets but also a trained agent. On the other hand, most of existing methods leave nothing behind except extracted itemsets. In other words, GIM-RL retains knowledge obtained in the mining process as the trained agent, while it is quite wasteful that most of existing methods throw such knowledge out. With respect to this, there exist many cases where the same mining process is performed on multiple related datasets. For example, one may want to inspect trend changes by extracting HUIs from two datasets collected in different terms. It is reasonable to use knowledge obtained from one of these datasets to speed up the mining process on the other dataset. Moreover, if the latter dataset is too huge to perform itemset mining from scratch, knowledge from the former dataset may be useful for accomplishing adequate mining on the latter one.

Based on the above consideration, we investigate GIM-RL's novel potential called \textit{agent transfer}. This means that an agent is firstly trained on a source dataset, and then it is transferred to another compatible agent for a target dataset. If the source and target datasets are related to each other, they are thought to be characterised by similar relations among items. Thus, the agent for the source dataset is expected to be useful also for the target dataset. That is, the transferred agent only needs fine-tuning  and offers more efficient itemset mining on the target dataset, compared to an agent trained from scratch. Unlike itemset mining, this kind of ``transfer learning'' is popular in different application domains where a model pre-trained on a source dataset is transferred to another model designed for a target dataset~\cite{M_Oquab,F_Li}. We bring transfer learning into itemset mining as agent transfer, and show its possibility to significantly accelerate itemset mining. 

This paper is organised as follows: The next section provides a survey of existing itemset mining methods to clarify the advantages of GIM-RL. Section~\ref{sec:gim-rl} presents a   methodological explanation of GIM-RL together with our reward designs to extract HUIs, FIs and ARs. The experimental results using GIM-RL based on these rewards are shown in Section~\ref{sec:exp}. Here, our agent transfer approach and the results demonstrating its effectiveness are also described. Section~\ref{sec:conc} concludes this paper by discussing several future directions to extend GIM-RL. In addition to the above-mentioned main contents, Appendixes~\ref{sec:app_pseudo} to \ref{sec:app_mushroom} give implementation details of GIM-RL, additional experiments and a small remark of one experimental dataset. Finally, many abbreviations and mathematical symbols are used in this paper. Thus, Appendix~\ref{sec:app_abb_sym} offers a list of abbreviations and the one of symbols in order for readers to follow this paper more easily.

\section{Related Work}
\label{sec:related}

Existing itemset mining methods are roughly divided into two categories, \textit{exhaustive} and \textit{non-exhaustive}. The former includes methods that enumerate all the itemsets matching a target type. However, the runtime of exhaustive methods significantly degrades as a dataset enlarges, especially, the increase in the number of distinct items causes the exponential expansion of the search space. To overcome this, non-exhaustive methods generate an approximate set of itemsets matching the target type. Below we first review several existing methods in the exhaustive and non-exhaustive categories, and then clarify the advantages of GIM-RL compared to those methods.

In general, an exhaustive method uses a data structure or algorithm specialised to a target type. For example, considering the downward closure property for FIs, Apriori algorithm~\cite{R_Agrawal} and FP-growth (Frequent Pattern growth) algorithm based on FP-tree~\cite{J_Han} are developed for efficient enumeration of FIs. Apart from FIs, a divide-and-conquer approach based on a bitwise vertical representation of a dataset is developed to extract closed FIs, each of which has no superset supported by the same set of transactions~\cite{C_Lucchese}. The method in \cite{T_Uno} enumerates maximal FIs that are included in no other FI, by devising a depth-first itemset expansion and dataset reductions. For efficient extraction of weighted FIs consisting of items associated with high weights, FP-growth is extended by crafting upper bound weights, three pruning techniques and a parallel mining algorithm~\cite{R_Kiran}. Infrequent weighted itemsets consisting of rare and lowly weighted items are extracted by revising FP-growth with a specialised interestingness measure and a technique of early discarding unpromising items~\cite{L_Gagliero} (please see \cite{Y_Koh} for a survey of existing infrequent itemset mining methods and \cite{S_Darrab} for an up-to-date survey). An HUI is an extension of a weighted FI in the sense that its utility is computed by considering both the weight and quantity of each item in a transaction. To efficiently extract all the HUIs, researchers have developed a list that facilitates the expansion of an itemset and the corresponding utility calculation~\cite{M_Liu}, and a tree structure based on the downward closure property of upper bound utilities~\cite{V_Tseng,C_Ahmed}. Finally, to extract FIs satisfying certain constraints like ``the median of weights of items in an FI must be larger than a threshold'', those constraints are converted so as to exhibit the downward closure property, and incorporated into FP-growth~\cite{J_Pei2}.

Exhaustive methods reviewed above are inflexible because a different data structure or algorithm is needed to extract a different type of itemsets. In contrast, GIM-RL can extract diverse types of itemsets only by defining a reward suitable for each type. By referring to the reward definitions for HUIs and FIs in Section~\ref{sec:reward}, one can easily design rewards for weighted FIs and infrequent weighted itemsets. In addition, it is expected that most of maximal FIs can be extracted using a reward, which is computed by checking the frequency of an itemset and the existence of its supersets in the set of already explored itemsets. This reward leads an agent to form itemsets that not only are frequent but also include more items. Closed FIs are likely to be extracted using a similar reward. Furthermore, many types of itemsets like ARs are divided into the antecedent and consequent, and let us assume the extraction of itemsets characterised by a consequent with one item. Under this setting, by exploiting the approach described in Section \ref{sec:reward_ar}, GIM-RL is expected to train an agent that can extract itemsets matching each of $38$ interestingness measures listed in \cite{L_Geng}. 

One of the most popular non-exhaustive itemset mining approaches is \textit{Evolutionary Computation} (EC) that offers a metaheuristic where nature-inspired operators are iteratively used to update itemsets into better ones in terms of a fitness function~\cite{A_Telikani}. One feature of EC-based methods is their predictable runtimes because itemset extraction is terminated by a specified number of iterations and no complicated process is needed for updating itemsets. One main class of EC-based methods is characterised by \textit{Genetic Algorithm} (GA) that iteratively selects promising itemsets using a fitness function, and exploits them to create new itemsets based on crossover and mutation operators~\cite{J_Mata,W_Song2}. Another main class is based on \textit{swarm intelligence-based algorithms} that iteratively update itemsets based on operations inspired by the collective behaviours of swarms like ants, bees and bats~\cite{W_Song,A_Telikani}.

Since there is no guarantee that GIM-RL can extract all the itemsets of a target type, it is classified as a non-exhaustive method and has a high similarity to EC-based methods. This is because a reward in GIM-RL corresponds to a fitness function in EC-based methods, and any type of itemsets can be extracted by defining a suitable fitness function for the type. However, the biggest difference is that each EC-based method relies on a heuristically pre-defined strategy (i.e., metaheuristic) to extract itemsets, while GIM-RL analyses a dataset and learns such a strategy as a trained agent. In other words, the former only outputs extracted itemsets, whereas GIM-RL produces those itemsets as well as the trained agent that captures generalised characteristics of items in the dataset and can be transferred for another similar dataset.

Another popular approach to non-exhaustive itemset mining is \textit{pattern sampling} that approximates a probability distribution over the search space by associating each itemset with a probability, which is proportional to its relevance to a target type in terms of an interestingness measure~\cite{M_Boley,V_Dzyuba}. Thus, itemsets sampled according to this probability distribution are a representative subset of itemsets matching the target type. However, only a limited number of itemset types can be treated by pattern sampling because of their compatibilities with sampling algorithms, for instance, an itemset type needs to be represented by a specified weight function form~\cite{M_Boley} or by a combination of XOR constraints~\cite{V_Dzyuba}. GIM-RL can extract a much more variety of itemsets. In addition, no consideration is given on whether a probability distribution approximated for a source dataset can be transferred to the one for a target dataset. For this, GIM-RL offers a very flexible agent transfer in which an agent can be transferred between the source and target datasets even if they are characterised by different sets of distinct items.

Agent transfer is related to incremental itemset mining where a dataset is updated by adding, deleting and modifying transctions~\cite{P_Viger_ASO,C_Ahmed,C_Leung}, and itemset mining in a stream where transactions arrive in rapid succession~\cite{P_Viger_ASO,T_Calders}. While these two tasks treat datasets that change over time, source and target datasets for agent transfer are fixed before extracting itemsets. As another substantial difference, the above two tasks focus on data structures to efficiently manage or summarise information necessary for extracting itemsets~\cite{C_Ahmed,C_Leung,T_Calders}. In agent transfer, such information is held by an agent (neural network) that captures generalised relations among items in the source or target dataset. 

Another related approach is multitask itemset mining that performs joint analysis of multiple related datasets to extract ``global'' itemsets which are eligible on most of those datasets~\cite{P_Taser}. Compared to this, agent transfer carries out ordered analysis of source and target datasets because an agent is firstly trained on the former and then transferred to the latter. In addition, the method in \cite{P_Taser} extracts global itemsets by directly holding and checking ``local'' itemsets extracted from each of multiple datasets. With respect to this, instead of itemsets extracted from the source dataset, agent transfer holds the trained agent as the abstracted information of those itemsets.

To our best knowledge, GIM-RL is the first method that adopts RL to train a machine learning model (i.e., agent defined by a neural network) for itemset mining. Correspondingly, agent transfer is not explored in any existing work. As discussed in Section \ref{sec:exp_runtime}, one problem of GIM-RL is its slow runtime because it needs multiple scans over a dataset to compute rewards and states of an environment. Of course, fast reward/state computation is one important future work. But, we believe that the efficiency of GIM-RL cannot be measured only by its runtime. The reason is that, according to a user-defined reward for a target type, GIM-RL adaptively trains an agent that can extract itemsets of this type, although one or more days may be needed. This seems much more efficient compared to a case where someone spends one or more months to implement a specialised data structure or algorithm for the target type.

\section{GIM-RL}
\label{sec:gim-rl}

In this section, we first describe GIM-RL's general framework that can be commonly used to extract various types of itemsets. Then, we provide specific reward definitions to extract HUIs, FIs and ARs. Finally, the agent training process of GIM-RL is intuitively illustrated using a small example dataset. Also, a list of mathematical symbols used to explain GIM-RL is provided in Appendix~\ref{sec:app_abb_sym} to make this section easier to follow.

\subsection{General Framework}
\label{sec:framework}

\begin{figure*}[htbp]
	\centering
	\includegraphics[width=\linewidth]{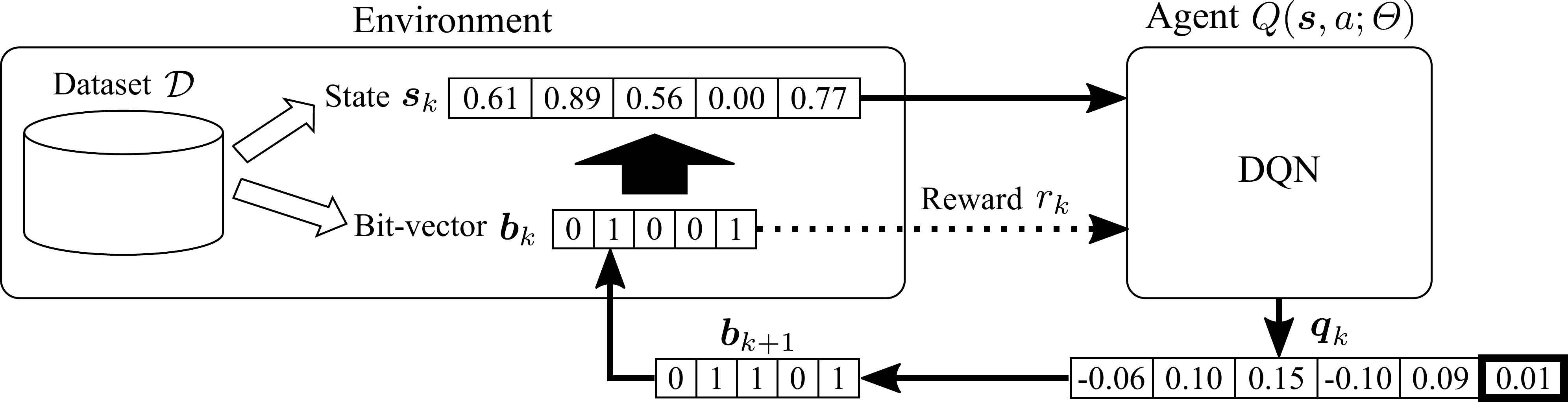}
	\caption{An overview of GIM-RL.}
	\label{fig:overview}
\end{figure*}

Let $\mathcal{D} = \{T_1, \cdots, T_N\}$ be a dataset containing $N$ transactions, and $\mathcal{I} = \{i_1,  \cdots, i_M\}$ be the set of $M$ distinct items each of which is included in at least one transaction in $\mathcal{D}$. Each transaction $T_n$ ($1 \leq n \leq N$) in $\mathcal{D}$ is a subset of $\mathcal{I}$ (i.e., $T_n \subseteq \mathcal{I}$). Letting $|T_n|$ denote the number of items in $T_n$, we describe $T_n = \{ i_{n,1}, \cdots , i_{n,|T_n|} \}$ where $i_{n,l}$ ($1 \leq l \leq |T_n|$) is the $l$th item in $T_n$ (i.e., $i_{n,l} \in \mathcal{I}$). We denote by $X$ an itemset consisting of items in $\mathcal{I}$ (i.e., $X \subseteq \mathcal{I}$). Let us consider an interestingness measure $\varphi (X)$ that takes $X$ as input, analyses transactions in $\mathcal{D}$ and outputs a value expressing the relevance of $X$ to a target type. For example, $\varphi(X)$ for FIs returns the support (frequency) of $X$ in $\mathcal{D}$ and $\varphi(X)$ for HUIs outputs $X$'s utility. In addition, $\varphi (X)$ can be flexibly used for an itemset type requiring multiple conditions. For instance, to extract ARs by considering the support and confidence of $X$, one can design $\varphi(X)$ that outputs a value depending on whether $X$ meets either or both of support and confidence thresholds. Under the setting described above, the goal of itemset mining is to extract from $\mathcal{D}$ every itemset $X$ for which $\varphi (X)$ is larger than a pre-defined threshold $\xi$ (i.e., $\varphi (X) \geq \xi$).

We formulate itemset mining as an RL problem shown in Fig.~\ref{fig:overview}. It is assumed that the environment signifies one step of the mining process. Specifically, the $k$th step is performed to modify the itemset defined by a bit-vector $\boldsymbol{b}_k$ into a new one that is likely to match a target type. Formally, $\boldsymbol{b}_k  = \left(b_{k,1}, \cdots, b_{k,M} \right)^T$ is an $M$-dimensional binary vector where $b_{k,m} \in \{0, 1\}$ ($1 \leq m \leq M$) represents the inclusion of the $m$th item $i_m$ in $\mathcal{I}$, namely, $b_{k,m} = 1$ indicates $i_m$ is included, otherwise not-included. We express the itemset defined by $\boldsymbol{b}_k$ as $X(\boldsymbol{b}_k)$. The environment produces a state $\boldsymbol{s}_k = \left(s_{k,1}, \cdots, s_{k,M} \right)^T$ having the same dimensionality to $\boldsymbol{b}_k$. Here, $s_{k,m}$  ($1 \leq m \leq M$) exhibits how useful it is to change $b_{k,m}$ for forming an itemset of the target type. In our implementation, $s_{k,m}$ is computed based on the simulation of the one-step-ahead future, in which $b_{k,m}$ is virtually changed to create the bit-vector $\boldsymbol{b}'_k$ and the corresponding itemset $X ( \boldsymbol{b}'_k )$. That is, $X ( \boldsymbol{b}_k )$ and $X ( \boldsymbol{b}'_k )$ differ only in the inclusion of $i_m$. Then, $s_{k,m}$ is calculated as $\varphi (X ( \boldsymbol{b}'_k ))$ by applying the interestingness measure $\varphi (X)$ to $X ( \boldsymbol{b}'_k )$. Please see Appendix~\ref{sec:app_dqn_arch} for more details to compute $s_{k,m}$. Since the computation of $\boldsymbol{s}_k$ needs to check transactions in $\mathcal{D}$, $\boldsymbol{s}_k$ in Fig.~\ref{fig:overview} is connected with the non-filled arrow from $\mathcal{D}$. 

$\boldsymbol{s}_k$ shows hints about which items are likely to be included or excluded to form an itemset of the target type. Thus, as illustrated by the solid arrow from $\boldsymbol{s}_k$ to the agent in Fig.~\ref{fig:overview}, $\boldsymbol{s}_k$ is fed into the agent to decide an action of changing one value in $\boldsymbol{b}_k$. As a result, $\boldsymbol{b}_k$ is updated into $\boldsymbol{b}_{k+1}$, meaning that the new itemset $X ( \boldsymbol{b}_{k+1} )$ is created by adding or removing one item to or from $X ( \boldsymbol{b}_k )$. The agent's action is designed based on human's itemset search where he/she modifies an itemset by thinking about which item's inclusion or exclusion is crucial for the target type. Then, the environment generates a reward $r_{k}$ that has a close  relation to $\varphi \left(X \left( \boldsymbol{b}_{k+1} \right) \right)$. Like $\boldsymbol{s}_k$, the computation of $r_k$ needs to check occurrences of $X ( \boldsymbol{b}_{k+1} )$ in $\mathcal{D}$, so a non-filled arrow is placed between the bit-vector (that is now $\boldsymbol{b}_{k+1}$) and $\mathcal{D}$ in Fig.~\ref{fig:overview}. As depicted by the dashed arrow in Fig.~\ref{fig:overview}, the agent receives $r_{k}$ as an evaluation score for the action taken at the $k$th step. If $X ( \boldsymbol{b}_{k+1} )$ matches the target type, the agent obtains high $r_{k}$. Afterwards, the environment creates a new state $\boldsymbol{s}_{k+1}$ based on which the agent updates $\boldsymbol{b}_{k+1}$ into $\boldsymbol{b}_{k+2}$ and receives $r_{k+1}$. This way, the agent iteratively updates the bit-vector to form various itemsets that possibly match the target type. In this framework, our goal is to train the agent that maximises cumulative rewards over steps. This means that the agent can extract as many itemsets matching the target type as possible.

It should be noted that $\boldsymbol{s}_k$ only provides the information for selecting one action, and does not include information for determining a sequence of multiple actions. In other words, the target type of itemsets are not necessarily extracted by greedily changing $\boldsymbol{b}_k$'s value that corresponds to $\boldsymbol{s}_k$'s highest value. Instead, the agent needs to have an intelligent policy to select an action at the current step by considering what kind of itemsets will be obtained at the future steps. Intuitively, we aim to train an agent that takes actions causing itemsets of non-target types at some consecutive steps, in order to extract many itemsets of the target type after those steps.

To this end, we employ \textit{Q-learning} that trains an agent characterised by a \textit{Q function} which takes as input an action $a$ and a state $\boldsymbol{s}$ of the environment, and outputs $a$'s quality at $\boldsymbol{s}$~\cite{V_Lavent,V_Mnih,V_Mnih2015}. Ideally, the optimal Q function $Q^* (\boldsymbol{s}, a)$ quantifies $a$'s quality as follows:
\begin{equation}
	Q^* (\boldsymbol{s}, a) = \max_{\pi} \mathbb{E} \left[ \sum_{k'=k}^{K} \gamma^{k'-k} r_{k'} \ \Big| \ \tilde{\boldsymbol{s}}_k = \boldsymbol{s}, \tilde{a}_k = a, \pi \right],
	\label{eq:q_optimal}
\end{equation}
where $K$ is the final step of the mining process, $\gamma$ is a discount factor by which a reward at a further future step is discounted more strongly, and $\pi$ represents a policy for action selection. Eq.~\ref{eq:q_optimal} means that $a$'s quality at $\boldsymbol{s}$ is computed as the maximum expected cumulative reward, which is achievable after seeing $\boldsymbol{s}$ at the $k$th step and taking $a$. Here, $\boldsymbol{s}$ and $a$ are assigned to the variables $\tilde{\boldsymbol{s}}_k$ and $\tilde{a}_k$ for representing a state and an action at the $k$th step, respectively. Simply speaking, in itemset mining, $Q^* (\boldsymbol{s}, a)$ indicates the maximum number of itemsets of the target type after taking $a$ at $\boldsymbol{s}$. However, it is impractical to directly build $Q^* (\boldsymbol{s}, a)$ because of the huge number of state-action combinations. Thus, considering the recent success of deep Q-learning~\cite{V_Lavent,V_Mnih,V_Mnih2015}, we approximate $Q^* (\boldsymbol{s}, a)$ by a neural network, called \textit{Deep Q-Netwrok} (DQN), defined by a set of parameters $\varTheta$. Therefore, $Q^* (\boldsymbol{s}, a)$ is  parametrised as $Q(\boldsymbol{s}, a; \varTheta)$, and our goal is to optimise $\varTheta$ of the DQN so that actions selected based on $Q(\boldsymbol{s}, a; \varTheta)$ form many itemsets of the target type. Note that we interchangeably use the terms ``agent'' and ``DQN'' and the notation ``$Q(\boldsymbol{s}, a; \varTheta)$'' in the following discussions.

Let $\mathcal{A}$ be a set of possible actions. In Fig.~\ref{fig:overview}, $Q(\boldsymbol{s}, a; \varTheta)$ is used to compute $Q(\boldsymbol{s}_k, a; \varTheta)$ that approximates the quality of each action $a \in \mathcal{A}$ at the specific state $\boldsymbol{s}_k$ of the $k$th step. Then, the action $a_k$ is chosen as the one corresponding to the highest value of $Q(\boldsymbol{s}_k, a; \varTheta)$. With no deep elaboration, one may define $\mathcal{A}$ to include $M$ actions each of which changes $b_{k,m}$ in $\boldsymbol{b}_k$ (i.e., $|\mathcal{A}| = M$). But, as depicted by the bold-lined rectangle at the bottom-right of Fig.~\ref{fig:overview}, in order to help the agent explore diverse itemsets, we add to $\mathcal{A}$ one more action that randomly initialises the bit-vector. This means that the agent is given an opportunity to stop updating $\boldsymbol{b}_k$ and re-start itemset mining from the initialised bit-vector. As an implementation detail, the random bit-vector initialisation is performed based on the probability distribution where the probability of $b_{k,m} = 1$ is proportional to the frequency of the $m$th item $i_m$ in $\mathcal{D}$, and is repeated until the bit-vector defines an itemset that exists in $\mathcal{D}$. In other words, it is meaningless to examine itemsets not-existing in $\mathcal{D}$. 

Also, an $(M+1)$-dimensional vector $\boldsymbol{q}_k = ( q_{k,1}, \cdots, q_{k,M},$\\$q_{k,M+1})^T$ is defined to represent the collection of $Q(\boldsymbol{s}_k, a; \varTheta)$s for all the $M+1$ actions at the $k$th step, as depicted at the bottom-right of Fig.~\ref{fig:overview}. That is, $q_{k,m}$ ($1 \leq m \leq M$) indicates an approximate quality of the action to change the inclusion of the $m$th item $i_m$ in $X ( \boldsymbol{b}_{k} )$, and $q_{k,M+1}$ expresses an approximate quality of the random bit-vector initialisation. $\boldsymbol{q}_k$ will be used to simplify the descriptions in Section~\ref{sec:exp_im} and Appendix~\ref{sec:app_dqn_arch}.

According to Bellman equation, the optimisation of $\varTheta$ is done by making $Q(\boldsymbol{s}_k, a_k; \varTheta)$ and $r_k + \gamma \max_{a'  \in \mathcal{A}} Q(\boldsymbol{s}_{k+1}, a'; \varTheta)$ as close as possible~\cite{V_Lavent,V_Mnih,V_Mnih2015}. An intuitive interpretation is that if $Q(\boldsymbol{s}, a; \varTheta)$ is a good approximation of $Q^* (\boldsymbol{s}, a)$, the maximum expected cumulative reward estimated for taking $a_k$ at $\boldsymbol{s}_k$ should be equal or very similar to the sum of $r_k$ obtained by taking $a_k$ at $\boldsymbol{s}_k$ with the maximum expected cumulative reward estimated for the action at the next state $\boldsymbol{s}_{k+1}$. However, the following two issues make it unstable to optimise $\varTheta$ by directly using $Q(\boldsymbol{s}_k, a_k; \varTheta)$ and $r_k + \gamma \max_{a' \in \mathcal{A}} Q(\boldsymbol{s}_{k+1}, a'; \varTheta)$. First, the agent's experiences $(\boldsymbol{s}_k,a_k,r_k,\boldsymbol{s}_{k+1})$s obtained at consecutive steps are very similar because only one value in $\boldsymbol{b}_k$ is changed at each step except its random initialisation. As a result, $Q(\boldsymbol{s}, a; \varTheta)$ trained on those experiences is very biased. To overcome this, we use a \textit{replay memory} $\mathcal{P}$ that is a queue to store the recent $|\mathcal{P}|$ experiences, and optimise $\varTheta$ on experiences randomly sampled from $\mathcal{P}$~\cite{V_Mnih,V_Mnih2015}. Second, $\varTheta$ is included in the ``target value $r_k + \gamma \max_{a' \in \mathcal{A}} Q(\boldsymbol{s}_{k+1}, a'; \varTheta)$'' that $Q(\boldsymbol{s}_k, a_k; \varTheta)$ targets to approximate. That is, the target value changes every time $\varTheta$ is updated, which makes the optimisation of $\varTheta$ unstable. To address this, separate DQNs are used for $Q(\boldsymbol{s}_k, a_k; \varTheta)$ and the target value~\cite{V_Mnih,V_Mnih2015}. The parameters $\varTheta$ of the DQN $Q(\boldsymbol{s}, a; \varTheta)$ are updated every step. On the other hand, the DQN $Q(\boldsymbol{s}, a; \varTheta^{-})$ used for the target value is called \textit{target network} and characterised by the parameters $\varTheta^{-}$ that are periodically updated every a certain number of steps. To sum up, as shown in the following equation, $Q(\boldsymbol{s}, a; \varTheta)$ is trained by minimising the squared difference between $Q(\boldsymbol{s}_k, a_k; \varTheta)$ and the target value $r_k + \gamma \max_{a' \in \mathcal{A}} Q(\boldsymbol{s}_{k+1}, a'; \varTheta^{-})$ on experiences randomly sampled from $\mathcal{P}$:
\begin{align}
	\mathbb{E}_{(\boldsymbol{s}_k,a_k,r_k,\boldsymbol{s}_{k+1}) \sim \mathcal{P}} \bigg[ \Big( r_k + \gamma \max_{a' \in \mathcal{A}} & Q(\boldsymbol{s}_{k+1}, a'; \varTheta^{-}) \nonumber \\
	& -  Q(\boldsymbol{s}_k, a_k; \varTheta) \Big)^2 \bigg]
	\label{eq:q_loss}
\end{align}

Finally, GIM-RL to train an agent $Q(\boldsymbol{s}, a; \varTheta)$ is summarised in Algorithm~\ref{alg:gim-rl} of Appendix~\ref{sec:app_pseudo}. Overall, GIM-RL executes $E$ episodes consisting of $K$ steps where the agent updates a bit-vector to form different itemsets. Here, each episode is used as a unit of itemset mining in the sense that a new mining process is initiated by a randomly initialised bit-vector, as shown at line 7 in Algorithm~\ref{alg:gim-rl} of Appendix~\ref{sec:app_pseudo}. Although itemset mining is not episodic, we believe that using episodes is useful for the agent to reinitialise the itemset search and explore diverse itemsets.

\subsection{Reward Definition}
\label{sec:reward}

We explain our definitions of rewards used to extract HUIs, FIs and ARs. Of course, it is possible to define better rewards than ours below. All the rewards in this paper are based on the common scheme dealing with the following four cases:

\noindent \textbf{Case 1:} The itemset $X(\boldsymbol{b}_{k+1})$ resulting from the action $a_k$ at the state $\boldsymbol{s}_k$ does not exist in $\mathcal{D}$. The agent receives a reward of $-1$, so that it is guided to not perform $a_k$ at $\boldsymbol{s}_k$ as well as at a similar state.

\noindent \textbf{Case 2:} The interestingness measure value $\varphi (X(\boldsymbol{b}_{k+1}))$ is less than the quarter of a pre-specified threshold $\xi$ (i.e., $\varphi (X(\boldsymbol{b}_{k+1})) < \xi / 4$) and the agent receives a reward of $0$. This means that taking $a_k$ at $\boldsymbol{s}_k$ or a similar state has no important impact on forming the target type of itemsets.

\noindent \textbf{Case 3:} $\varphi (X(\boldsymbol{b}_{k+1})) \geq \xi/4$ is the condition of this case, which is further divided into four sub-cases, $\xi/4 \leq \varphi (X(\boldsymbol{b}_{k+1})) < \xi/2$, \ $\xi/2 \leq \varphi (X(\boldsymbol{b}_{k+1})) < 3\xi/4$, \ $3\xi/4 \leq \varphi (X(\boldsymbol{b}_{k+1})) < \xi$ and $\xi \leq \varphi (X(\boldsymbol{b}_{k+1}))$, in which the agent receives rewards of $1$, $2$, $3$ and $4$, respectively. These rewards are defined according to how close $X(\boldsymbol{b}_{k+1})$ is to the target type. Thereby, the agent is informed of how important it is to take $a_k$ at $\boldsymbol{s}_k$ or a similar state for forming itemsets of the target type. The reason why the last sub-case $\xi \leq \varphi (X(\boldsymbol{b}_{k+1}))$ is included is explained below. 

\noindent \textbf{Case 4:} This case is defined by two conditions. The first examines whether  $X(\boldsymbol{b}_{k+1})$ matches the target type (i.e., $\xi \leq \varphi (X(\boldsymbol{b}_{k+1}))$), and the second checks whether $X(\boldsymbol{b}_{k+1})$ has not yet been extracted in the current episode. $X(\boldsymbol{b}_{k+1})$ satisfying these two conditions is a newly extracted itemset of the target type, and we aim to extract such itemsets. Thus, by providing a very high reward of $100$, the agent is led to extract many unique itemsets of the target type. Also, if $X(\boldsymbol{b}_{k+1})$ is an already extracted itemset, it seems unreasonable to regard $a_k$ as meaningless because $X(\boldsymbol{b}_{k+1})$ anyway matches the target type. This situation is captured by the last sub-case in Case 3, and the agent gets a reward ($4$) that is much smaller than the one ($100$) in this case.

To complete the above-mentioned scheme, we describe the computation of $\varphi (X(\boldsymbol{b}_{k+1}))$ for each of HUI, FI and AR.

\subsubsection{$\varphi (X(\boldsymbol{b}_{k+1}))$ for HUI Extraction}
\label{sec:reward_hui}

For the $n$th transaction $T_n = \{i_{n,1}, \cdots, i_{n,|T_n|} \}$ in $\mathcal{D}$, the $l$th item $i_{n,l}$ ($1 \leq l \leq |T_n|$) is associated with the item-specific utility $p(i_{n,l})$ and the quantity $q(i_{n,l})$ in $T_n$. Under this setting, $\varphi (X(\boldsymbol{b}_{k+1}))$ representing the utility of $X(\boldsymbol{b}_{k+1})$ is computed as follows~\cite{V_Tseng,C_Ahmed,M_Liu,W_Song}:
\begin{align}
	\varphi & (X(\boldsymbol{b}_{k+1})) \nonumber \\
	& = \sum_{X(\boldsymbol{b}_{k+1}) \subseteq T_n \land T_n \in \mathcal{D}}  \ \sum_{i_{n,l} \in X(\boldsymbol{b}_{k+1}) \land i_{n,l} \in T_n} p(i_{n,l}) \ q(i_{n,l})
	\label{eq:measure_hui}
\end{align}
The inner summation refers to each transaction containing $X(\boldsymbol{b}_{k+1})$ as $T_n$, and computes the sum of item-specific utilities weighted by their corresponding quantities for all items in $X(\boldsymbol{b}_{k+1})$. As signified by the outer summation, $\varphi (X(\boldsymbol{b}_{k+1}))$ is calculated as the total of such weighted sums for all transactions containing $X(\boldsymbol{b}_{k+1})$.

\subsubsection{$\varphi (X(\boldsymbol{b}_{k+1}))$ for FI Extraction}
\label{sec:reward_fi}

$\varphi (X(\boldsymbol{b}_{k+1}))$ is defined as $sup(X(\boldsymbol{b}_{k+1}))$ representing the support (frequency) of $X(\boldsymbol{b}_{k+1})$ in $\mathcal{D}$, that is, the number of transactions containing $X(\boldsymbol{b}_{k+1})$.

\subsubsection{$\varphi (X(\boldsymbol{b}_{k+1}))$ for AR Extraction}
\label{sec:reward_ar}

\begin{figure*}[htbp]
	\centering
	\includegraphics[width=\linewidth]{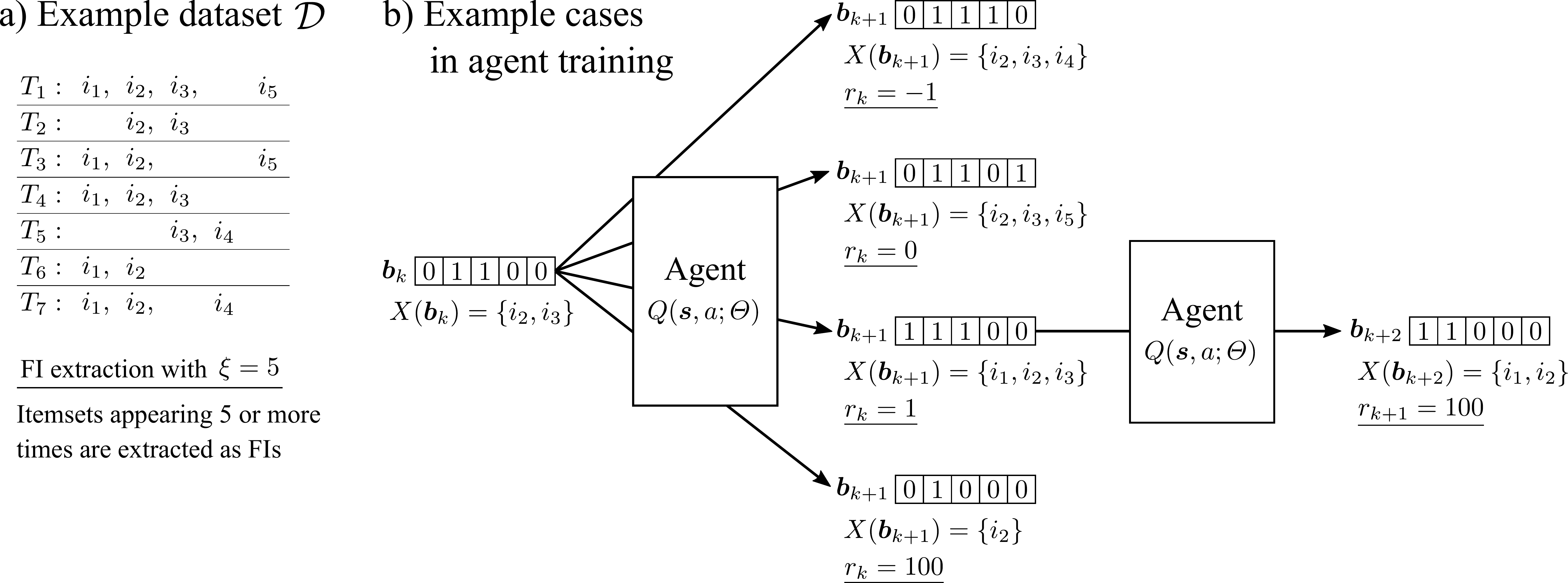}
	\caption{An illustration of how an agent $Q(\boldsymbol{s}, a; \varTheta)$ for FI extraction is trained in GIM-RL.}
	\label{fig:ill_ex}
\end{figure*}

To define an AR, an itemset $X$ is firstly divided into two subsets $X_a$ and $X_c$ (i.e., $X = X_a \cup X_c$), and a rule $X_a \rightarrow X_c$ is created by viewing $X_a$ and $X_c$ as the antecedent and consequent, respectively. $X_a \rightarrow X_c$ is regarded as an AR if $sup(X)$ is larger than the minimum support threshold $min\_sup$ and the confidence, which is the conditional probability of $X_c$ given $X_a$ (i.e., $sup(X)/sup(X_a)$), is larger than the minimum confidence threshold $min\_conf$. GIM-RL extracts ARs using $\varphi (X(\boldsymbol{b}_{k+1}))$ that jointly considers the support and confidence conditions.

As one remark, the current GIM-RL is limited to extracting ARs each of which is characterised by $X_c$ with one item. But, focusing only on such ARs is considered reasonable because allowing $X_c$ to include multiple items often produces an extremely large number of ARs. In addition, several existing methods address the extraction of ARs with $X_c$s characterised by single items~\cite{G_Webb,S_Surampudi}. Moreover, assuming that human is paying attention to one itemset, he/she is likely to think whether a rule is created by regarding the itemset as $X_a$ and a new item as $X_c$, or by seeing an item in the itemset as $X_c$ and the other items as $X_a$. GIM-RL implements this human's search based on the difference between $X(\boldsymbol{b}_k)$ and $X(\boldsymbol{b}_{k+1})$. If $X(\boldsymbol{b}_{k+1})$ is the itemset created by adding one item to $X(\boldsymbol{b}_k)$, we consider that the agent takes the action to use $X(\boldsymbol{b}_k)$ and the added item as $X_a$ and $X_c$, respectively. For an inverse case where $X(\boldsymbol{b}_{k+1})$ is the itemset created by removing one item from $X(\boldsymbol{b}_k)$, the agent's action is to use $X(\boldsymbol{b}_{k+1})$ and the removed item as $X_a$ and $X_c$, respectively.

Such an agent's action to decide $X_a$ and $X_c$ that may constitute an AR, is evaluated by a reward based on two interesting measures $\varphi_1 (X(\boldsymbol{b}_{k+1}))$ and $\varphi_2 (X(\boldsymbol{b}_{k+1}))$. The former is used in Cases 2 and 3 to check the support of $X_a \cup X_c$ which is the itemset including more items between $X(\boldsymbol{b}_k)$ and $X(\boldsymbol{b}_{k+1})$. For Case 3 where $\varphi_1 (X(\boldsymbol{b}_{k+1})) = sup(X_a \cup X_c)$ is equal to or larger than $\xi_1 / 4 = min\_sup  / 4$, the agent gets a reward of $1$, $2$, $3$ or $4$. This is because these $X_a$ and $X_c$ are close to constituting an AR in terms of the support condition. $\varphi_2 (X(\boldsymbol{b}_{k+1}))$ is used in Case 4 and checks the confidence of $X_a \rightarrow X_c$, that is, $\varphi_2 (X(\boldsymbol{b}_{k+1})) = sup(X_a \cup X_c)  / sup(X_a)$. Note that Case 4 is triggered only when Case 3 is passed by $min\_sup \leq \varphi_1 (X(\boldsymbol{b}_{k+1}))$. Thus, if $\xi_2 = min\_conf \leq \varphi_2 (X(\boldsymbol{b}_{k+1}))$, $X_a$ and $X_c$ obtained by the agent action are verified as the AR $X_a \rightarrow X_c$ and a reward of $100$ is given. Otherwise, the agent gets a reward of $4$ resulting from Case 3.

Finally, it is possible to extract ARs that are individually characterised by $X_c$ with multiple items, by allowing an agent to take actions for changing multiple values in the bit-vector. However, this causes an exponential increase of the agent's action space, so a smart approach is needed. This issue is left as our future work.

\subsection{Illustrative Example}
\label{sec:ill_ex}

For clear understanding of GIM-RL, we explain an intuitive example of how an agent $Q(\boldsymbol{s}, a; \varTheta)$ is trained by referring to Fig.~\ref{fig:ill_ex}. First of all, let us assume that an agent for FI extraction is trained using the dataset $\mathcal{D}$ shown in Fig.~\ref{fig:ill_ex} (a). Here, $\mathcal{D}$ contains $N=7$ transactions $T_1, \cdots, T_7$ defined on $M=5$ distinct items $i_1, \cdots, i_5$. The minimum support threshold $\xi = min\_sup$ is set to five. Hence, the agent is trained by optimising its parameters $\varTheta$ to extract itemsets appearing five or more times in $\mathcal{D}$ as FIs.

Fig.~\ref{fig:ill_ex} (b) depicts some example cases that can possibly occur at the $k$th, $(k+1)$th and $(k+2)$th steps to train an agent on $\mathcal{D}$. The centre of Fig.~\ref{fig:ill_ex} (b) shows four cases that originate from the bit-vector at the $k$th step $\boldsymbol{b}_{k} = (0,1,1,0,0)^T$ defining the itemset $X(\boldsymbol{b}_{k}) = \{i_2,i_3\}$. The case at the top is characterised by $\boldsymbol{b}_{k+1} = (0,1,1,1,0)^T$ resulting from the action to change $i_4$'s inclusion in $X(\boldsymbol{b}_{k})$. According to our reward definition, the reward $r_k = -1$ is given to the agent because  $X(\boldsymbol{b}_{k+1}) = \{i_2, i_3, i_4\}$ defined by $\boldsymbol{b}_{k+1}$ does not exist in $\mathcal{D}$. As a result, $\varTheta$ is updated so that $Q(\boldsymbol{s}, a; \varTheta)$ outputs a very small value for the action to change $i_4$'s inclusion at the state $\boldsymbol{s}_k$ computed from $\boldsymbol{b}_k$. To put it more simply, the agent becomes to not select this action for $\boldsymbol{b}_k$.

It should be noted that the above-mentioned update of $\varTheta$ affects action selection also for a bit-vector which is ``statistically similar'' to $\boldsymbol{b}_k$, although illustrating this in Fig.~\ref{fig:ill_ex} is difficult. The key is that $Q(\boldsymbol{s}, a; \varTheta)$ is defined on a state $\boldsymbol{s}$ represented by a continuous vector. Regarding this, two bit-vectors are considered statistically similar if items contained in the itemsets defined by them have similar statistical characteristics. Accordingly, the states computed from these bit-vectors are characterised by similar continuous vectors, for which $Q(\boldsymbol{s}, a; \varTheta)$ outputs similar values. Thus, when $\varTheta$ is updated to select (or not select) a certain action at a particular state, this action selection is propagated to states computed from statistically similar bit-vectors. In what follows, such bit-vectors are not described for the sake of brevity. But, please note that $\varTheta$'s updates described below influence on action selection for statistically similar bit-vectors.

The cases shown at the second-top, third-top and bottom of the centre of Fig.~\ref{fig:ill_ex} (b) can be interpreted in a similar way to the case at the top. The case at the second-top is based on $\boldsymbol{b}_{k+1} = (0,1,1,0,1)^T$ obtained by the action to change $i_5$'s inclusion in $X(\boldsymbol{b}_k)$. Since $X(\boldsymbol{b}_{k+1}) = \{i_2, i_3, i_5\}$ appears once in $\mathcal{D}$, the agent receives the reward $r_k = 0$. This means to postpone the evaluation of the action to change $i_5$'s inclusion at the state $\boldsymbol{s}_k$ computed from $\boldsymbol{b}_k$. Because of $r_k=0$, Eq.~\ref{eq:q_loss} requires $Q(\boldsymbol{s}_k, a_k; \varTheta)$ to be close to $\gamma \max_{a' \in \mathcal{A}} Q(\boldsymbol{s}_{k+1}, a'; \varTheta^{-})$. That is, $Q(\boldsymbol{s}, a; \varTheta)$ for the action to change $i_5$'s inclusion in $X(\boldsymbol{b}_k)$ depends on what rewards will be obtained at the $(k+1)$th and later steps. For the case at the third-top, changing $i_1$'s inclusion in $X(\boldsymbol{b}_{k})$ produces $X(\boldsymbol{b}_{k+1}) = \{i_1, i_2, i_3\}$ appearing twice in $\mathcal{D}$. Since this case is categorised as the first sub-case of Case 3 in our reward definition (i.e., $ \xi / 4 < sup(X(\boldsymbol{b}_{k+1})) = 2 < \xi / 2$), the agent receives the reward $r_k = 1$. As a result, $Q(\boldsymbol{s}, a; \varTheta)$ is refined to output a relatively high value for the action to change $i_1$'s inclusion at $\boldsymbol{s}_k$, making the agent possibly select this action. The case at the bottom is triggered by the action to change $i_3$'s inclusion in $X(\boldsymbol{b}_k)$ and the resulting $X(\boldsymbol{b}_{k+1}) = \{i_2\}$ appears six times in $\mathcal{D}$ and is extracted as an FI. The reward $r_k = 100$ guides $Q(\boldsymbol{s}, a; \varTheta)$ to output a high value for the action to change $i_3$'s inclusion at $\boldsymbol{s}_k$, making the agent likely to select this action.

One important remark is that $Q(\boldsymbol{s}, a; \varTheta)$ approximates the maximum expected cumulative reward defined in Eq.~\ref{eq:q_optimal}. Thus, training of $Q(\boldsymbol{s}, a; \varTheta)$ considers not only the reward $r_k$ at the $k$th step, but also rewards at future steps. In fact, as seen from Eq.~\ref{eq:q_loss}, $Q(\boldsymbol{s}, a; \varTheta)$ is not trained to directly approximate $r_k$, but to achieve a situation where the maximum expected cumulative reward approximated at the $k$th step is close to the sum of the actual reward $r_k$ at the $k$th step and the maximum expected cumulative reward approximated at the $(k+1)$th step\footnote{This objective progresses training of $Q(\boldsymbol{s}, a; \varTheta)$. The reason is that thanks to the consideration of the actual reward $r_k$, the target value $r_k + \gamma \max_{a' \in \mathcal{A}} Q(\boldsymbol{s}_{k+1}, a'; \varTheta^{-})$ is usually a more accurate approximation of the maximum expected cumulative reward than $Q(\boldsymbol{s}_k, a_k; \varTheta)$. Hence, the approximation by $Q(\boldsymbol{s}_k, a_k; \varTheta)$ can be more accurate by reducing the difference between it and the target value.}. Thereby, $Q(\boldsymbol{s}, a; \varTheta)$ reflects the estimation of what rewards will be acquired at future steps after taking $a$ at $\boldsymbol{s}$. This is illustrated in the case shown at the right of Fig.~\ref{fig:ill_ex} (b). This case is provoked by the action to change $i_3$'s inclusion in $X(\boldsymbol{b}_{k+1})$ and the resulting $X(\boldsymbol{b}_{k+2}) = \{i_1, i_2\}$ appearing five times in $\mathcal{D}$ is extracted as an FI. The reward $r_{k+1}=100$ obviously supervises $Q(\boldsymbol{s}, a; \varTheta)$ to output a high value for the action to change $i_3$'s inclusion at $\boldsymbol{s}_{k+1}$ computed from $\boldsymbol{b}_{k+1}$. In addition to this, $Q(\boldsymbol{s}, a; \varTheta)$ for the action to change $i_1$'s inclusion at $\boldsymbol{s}_{k}$ is also increased, because this action is now linked to the FI extracted at the $(k+2)$th step. This way, in GIM-RL, an agent is trained to select an action that leads to not only the immediate extraction of an FI at the current step but also the extraction of many FIs at future steps.

\section{Experimental Results}
\label{sec:exp}

We test GIM-RL using the five datasets shown in Table~\ref{tbl:datasets_stats}. These datasets are chosen because they have different characteristics in terms of numbers of transactions, numbers of distinct items and average numbers of items in one transaction, as exhibited in rows $N$, $M$ and ``Avg. $|T_n|$'' in Table~\ref{tbl:datasets_stats}, respectively. In addition, the datasets in Table~\ref{tbl:datasets_stats} are popularly used in many existing works~\cite{W_Song,W_Song2,M_Boley,V_Dzyuba,P_Viger_MTA}. Through the extractions of HUIs, FIs and ARs from these datasets, we aim to demonstrate the generality of GIM-RL.

\begin{table}[htbp]
	\caption{Statistics about the experimental datasets. }
	\centering
	\begin{tabular}{|r||c|c|c|c|c|} \hline
		& Chess    & Mushroom & Accidents\_$10$\%  & Connect   & Pumsb \\ \hline \hline
		$N$                      & $3196$ & $8416$      & $34018$  & $67557$ & $49046$ \\ \hline
		$M$                      & $75$      & $119$        & $468$       & $129$     & $2113$ \\ \hline
		Avg.                      & \multirow{2}{*}{$37$}  & \multirow{2}{*}{$23$}  & \multirow{2}{*}{$34$}   & \multirow{2}{*}{$43$}   & \multirow{2}{*}{$74$}  \\
		$|T_n|$               &                &                    &                    &                  &           \\ \hline
	\end{tabular}
	\label{tbl:datasets_stats}
\end{table}

\subsection{Results for Itemset Mining}
\label{sec:exp_im}

For each of the HUI, FI and AR extractions, the following two evaluations are performed: First, we use a baseline method that is developed to exhaustively enumerate all the itemsets of a target type. The number of itemsets extracted by the baseline method is maximum, so we examine how close the number of itemsets extracted by GIM-RL is to this maximum. Second, the effectiveness of agents trained by GIM-RL is examined by comparing the five agents below. Note that these agents commonly take $\boldsymbol{s}_k$ as input and output $a_k$ to change the value of one dimension in $\boldsymbol{b}_k$ in Fig.~\ref{fig:overview}, but they are different in how to decide $a_k$ based on $\boldsymbol{s}_k$.  

\noindent \textit{Ramdom:} This agent just changes the value of a randomly selected dimension in $\boldsymbol{b}_k$. With \textit{Random}, we aim to show the difficulty of each itemset extraction task where randomly changing values in  $\boldsymbol{b}_k$ yields few itemsets of the target type. 

\noindent \textit{State-}$\epsilon$\textit{:} As described in Section~\ref{sec:framework}, each dimension's value in $\boldsymbol{s}_k$ represents the usefulness of changing the value of the corresponding dimension in $\boldsymbol{b}_k$ based on the simulation of the one-step-ahead future. One may think that $\boldsymbol{s}_k$ already contains enough information about which value in $\boldsymbol{b}_k$ should be changed. To answer this, we test \textit{State-}$\epsilon$ that changes the value of $\boldsymbol{b}_k$'s dimension corresponding to the highest value in $\boldsymbol{s}_k$ with the probability $1 - \epsilon$, while changing the value of a randomly selected dimension with the probability $\epsilon$ to preserve the variety of itemsets to be explored.

\noindent \textit{State-prob:} The motivation for testing this agent is the same to the one for \textit{State-}$\epsilon$. But, different from \textit{State-}$\epsilon$, \textit{State-prob} determines $a_k$ by random sampling according to the probability distribution, where the probability of changing the value of one  dimension in $\boldsymbol{b}_k$ is proportional to the value of the corresponding dimension in $\boldsymbol{s}_k$.

\noindent \textit{GIM-RL-Basic:} This is the basic version of GIM-RL. A DQN that accepts $\boldsymbol{s}_k$ and outputs $\boldsymbol{q}_k$ is trained in the framework of Fig.~\ref{fig:overview}. Each dimension in $\boldsymbol{q}_k$ indicates an approximate quality of taking one of $M+1$ actions including the random bit-vector initialisation, as described in Section~\ref{sec:framework}. DQNs with the same architecture are trained for extracting HUIs and ARs, while simpler DQNs are used for FIs. Please see Appendix~\ref{sec:app_dqn_arch} (especially Fig.~\ref{fig:dqn_arch}) for details on the architectures of these DQNs.

\noindent \textit{GIM-RL-Fusion:} This is an extended version of \textit{GIM-RL-Basic} by defining the output as the sum of $\boldsymbol{q}_k$ and $\boldsymbol{s}_k$. To be precise, $\boldsymbol{s}_k$ misses the dimension for the random bit-vector initialisation, so $\boldsymbol{s}'_k$ is created by appending to $\boldsymbol{s}_k$ one dimension with the very small value ``$0.02 \times (\mbox{average of } \boldsymbol{s}_k)$'' for the initialisation. Then, the output of \textit{GIM-RL-Fusion} is computed as $\boldsymbol{q}'_k =  \lambda \boldsymbol{s}'_k + (1 - \lambda) \boldsymbol{q}_k$. Of course, the additional use of $\boldsymbol{s}'_k$ is expected to make $\boldsymbol{q}'_k$ attain better action selection than $\boldsymbol{q}_k$. But, $\boldsymbol{q}'_k$ plays a more important role to advance training of an agent. In principle, the agent can be trained when positive rewards are obtained. In other words, training of the agent does not proceed as long as rewards are zero or negative. Regarding this, at the beginning of training, it is statistically difficult for the agent to find ``positive itemsets'' that offer positive rewards using the imperfect $Q (\boldsymbol{s}, a; \varTheta)$. For this, $\boldsymbol{s}'_k$ significantly increases the probability of finding positive itemsets because it represents rough estimation of which items are likely to constitute positive itemsets. After finding some positive itemsets, the agent can be trained to some extend. This boosts the probability that the agent can find positive itemsets by its own exploitation of the trained $Q (\boldsymbol{s}, a; \varTheta)$, that is, the agent can do further training by itself. Considering this, $\lambda$ in $\boldsymbol{q}'_k$ is gradually reduced to weaken the effect of $\boldsymbol{s}'_k$ as training proceeds. Please see Appendix~\ref{sec:app_hyper_param} for the specific setting of $\lambda$.

\noindent \textbf{Results for HUI Mining:} Table~\ref{tbl:hui_result} summarises the results for HUI mining. First, HUI-Miner~\cite{M_Liu} implemented in SPMF library~\cite{P_Viger_SPMF} is used as a baseline to extract all the HUIs from each dataset. The thresholds in the second row of Table~\ref{tbl:hui_result} are the same to the ones used in \cite{W_Song,W_Song2}\footnote{Also for the other threshold settings used in \cite{W_Song,W_Song2}, we have obtained results to validate the effectiveness of GIM-RL.}. As shown in Table~\ref{tbl:hui_result}, $100$\% or nearly $100$\% of HUIs are extracted from each of the four datasets using \textit{GIM-RL-Fusion}. This verifies the effectiveness of GIM-RL to train agents for extracting HUIs. Also, the fact that no HUI is extracted by \textit{Random} implies the difficulty of HUI extraction. In addition, the poor performances of \textit{State-}$\epsilon$ and \textit{State-prob} indicate the ineffectiveness of directly using $\boldsymbol{s}_k$. In what follows, \textit{Random}, \textit{State-}$\epsilon$ and \textit{State-prob} are sometimes called ``non-training agents'' because their action selection policies are fixed in advance and are not optimised. Finally, the main reason why no HUI is extracted by \textit{GIM-RL-Basic} for Accidents\_$10$\% is that it fails to find first some positive itemsets, so its training does not proceed. In contrast, \textit{GIM-RL-Fusion} aided by $\boldsymbol{s}'_k$ stably produces very good performances on all the datasets.

\begin{table}[htbp]
	\caption{An overview of the HUI mining results.}
	\centering
	\begin{tabular}{|r||c|c|c|c|} \hline
		& \multirow{2}{*}{Chess}  & \multirow{2}{*}{Mushroom} & {\footnotesize Accidents}  & \multirow{2}{*}{Connect} \\ 
		&           &  & {\footnotesize \_$10$\%}  &  \\ \hline \hline
		$\xi$ (\%) & $29.0$\%	& $14.5$\%	 & $13.0$\%	& $32.0$\% \\ \hline
		Baseline     & $176$        & $199$          & $127$         & $171$ \\ \hline
		\textit{Random}                & $0$        & $0$          & $0$         & $0$ \\ \hline
		\textit{State-}$\epsilon$ & $56$     & $82$         & $51$      & $67$ \\ \hline
		\textit{State-prob}            & $0$        & $0$          & $0$         & $0$ \\ \hline
		\textit{GIM-RL-Basic}               & $176$   & $199$      & $0$        & $169$ \\ \hline
		\textit{GIM-RL-Fusion}             & $176$   & $197$     & $127$    & $169$ \\
		            & $(100\%)$ & $(98.9\%)$  & $(100\%)$  & $(98.8\%)$ \\ \hline
	\end{tabular}
	\label{tbl:hui_result}
\end{table}

\noindent \textbf{Results for FI Mining:} Table~\ref{tbl:fi_result} shows the results for FI mining. FP-growth~\cite{J_Han} implemented in SPMF library~\cite{P_Viger_SPMF} is selected as a baseline to extract all the FIs. By referring to \cite{P_Viger_MTA}, the thresholds in the second row are determined so that moderate numbers of FIs ($1000$-$10000$ FIs) are extracted. Note that FIs are extracted by the non-training agents, although numbers of extracted FIs significantly vary depending on datasets. For this, we point out the following two reasons attributed to the random bit-vector initialisation. First, a bit-vector is initialised according to the probability distribution based on the frequency of each item, so the itemset defined by this initialised bit-vector has a relatively high probability to be an FI. In addition, updating this bit-vector leads to find other FIs with not-low probabilities. Second, the initialisation is repeated until the bit-vector defines an itemset that exists in $\mathcal{D}$. Focusing only on such itemsets dramatically reduces the search space, especially for small datasets like Chess. Because of the above two reasons, even \textit{Random} can extract FIs. But, despite the fact that the non-training agents occasionally extract many FIs, \textit{GIM-RL-Fusion} or \textit{GIM-RL-Basic} extracts the highest numbers of FIs for all the datasets. This validates the effectiveness of GIM-RL also for extracting FIs\footnote{For Chess, $7901$ FIs ($96.0$\% of FIs) are extracted by \textit{GIM-RL-Fusion} using a DQN with the architecture for the HUI and AR extractions.}.

\begin{table}[htbp]
	\caption{An overview of the FI mining results.}
	\begin{tabular}{|r||c|c|c|c|} \hline
		& Chess          & Mushroom & Pumsb  & Connect \\ \hline \hline
		$\xi$ (\%)          & $80$\%	  & $35$\%	   & $90$\%	& $95$\% \\ \hline
		Baseline              & $8227$      & $1121$     & $2607$ & $2201$ \\ \hline
		\textit{Random}                 & $2258$ & $63$        & $304$   &$1541$ \\ \hline
		\textit{State-}$\epsilon$   & $587$   & $440$      & $427$   & $394$ \\ \hline
		\textit{State-prob}             & $4926$ & $502$     & $955$    & $2144$ \\ \hline
		\textit{GIM-RL-Basic}                & $6154$	& $1101$   & $2544$   & $2072$ \\ \hline
		\textit{GIM-RL-Fusion}              & $6843$ & $969$     & $2578$  & $2199$ \\
		& $(83.2\%)$ & $(86.4\%)$  & $(98.8\%)$  & $(99.9\%)$ \\ \hline \hline
	\end{tabular}
	\label{tbl:fi_result}
\end{table}

\noindent \textbf{Results for AR Mining:} In Table~\ref{tbl:ar_result}, the results for AR mining are presented. Our baseline method first uses FPGrowth\_association\_rules implemented in SPMF library~\cite{P_Viger_SPMF} to extract all the ARs, and then retains ARs each of which is characterised by a consequent with one item. The thresholds $\xi_1= min\_sup$ and $\xi_2 = min\_conf$ are set based on \cite{P_Viger_MTA}, so that a reasonable number of ARs are extracted from each dataset. The trend of the results in Table~\ref{tbl:ar_result} is similar to the one in Table~\ref{tbl:fi_result}. Because of the random bit-vector initialisation, the non-training agents can extract ARs from each dataset, but their performances significantly degrade on large datasets like Pumsb and Connect. In contrast, \textit{GIM-RL-Fusion} can stably extract almost all ARs from each of the datasets.

\begin{table}[htbp]
	\caption{An overview of the AR mining results.}
	\begin{tabular}{|r||c|c|c|c|} \hline
		& Chess          & Mushroom & Pumsb    & Connect \\ \hline \hline
		$\xi_1$ (\%)       & $90$\%      & $50$\%      & $90$\%  & $95$\% \\ \hline
		$\xi_2$ (\%)       & $80$\%      & $80$\%      & $80$\%  & $80$\%  \\ \hline
		Baseline               & $2351$      & $331$       & $11366$ & $10106$ \\ \hline
		\textit{Random}                 & $2214$ & $330$     & $120$      & $1287$ \\ \hline
		\textit{State-}$\epsilon$   & $639$   & $115$       & $640$    & $615$ \\ \hline
		\textit{State-prob}             & $2350$ & $331$    & $481$    & $3545$ \\ \hline
		\textit{GIM-RL-Basic}                & $2202$ & $331$     & $4832$  & $4026$ \\ \hline
		\textit{GIM-RL-Fusion}             & $2340$ & $330$     & $10111$  & $9225$ \\
		& $(99.5\%)$ & $(99.6\%)$  & $(88.9\%)$  & $(91.2\%)$ \\ \hline \hline
	\end{tabular}
	\label{tbl:ar_result}
\end{table}

\begin{figure}[htbp]
	\centering
	\includegraphics[width=\linewidth]{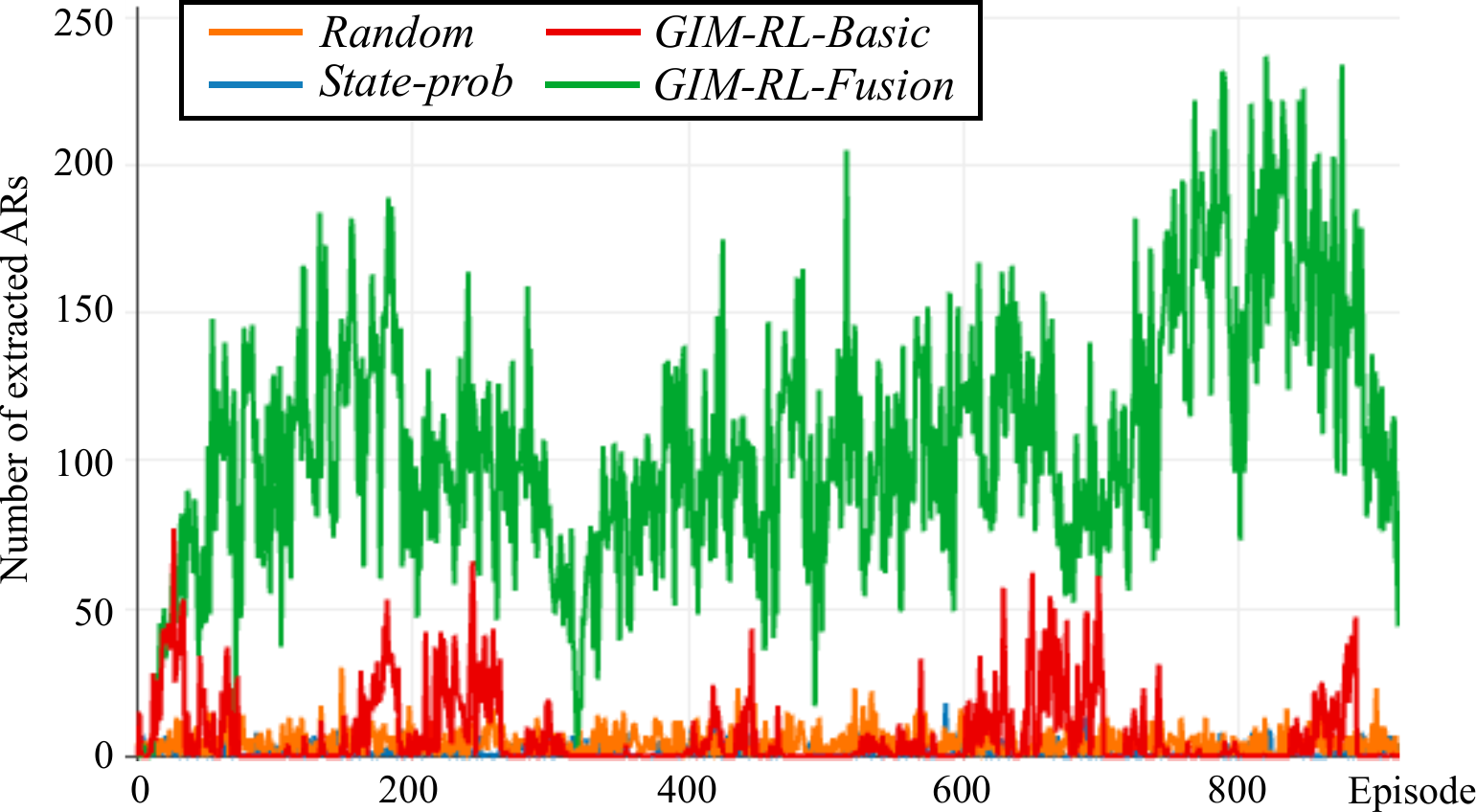}
	\caption{The transition of numbers of ARs extracted by each of \textit{Random}, \textit{State-prob}, \textit{GIM-RL-Basic} and \textit{GIM-RL-Fusion} over the passage of episodes (the dataset is Connect).}
	\label{fig:ar_episodes}
\end{figure}

We describe a deeper insight in the AR extraction on Connect. Fig.~\ref{fig:ar_episodes} shows how many ARs are extracted in each episode by \textit{Random}, \textit{State-prob}, \textit{GIM-RL-Basic} and \textit{GIM-RL-Fusion}. \textit{State-}$\epsilon$ is omitted because of the very low number of extracted ARs. In Fig.~\ref{fig:ar_episodes}, for each agent, the number of extracted ARs in one episode is counted without considering whether each AR is already extracted or not. Thus, the sum of numbers of ARs extracted by the agent over all the episodes is the total cumulative number of ARs. As shown in Fig.~\ref{fig:ar_episodes}, the numbers of ARs extracted by the non-training agents are constantly low over episodes, and they are nearly invisible. On the other hand, \textit{GIM-RL-Basic} is trained to extract ARs during about the first $50$ episodes, and afterwards keeps extracting ARs although the extraction is unstable as illustrated by the significantly varied numbers of ARs over episodes. Compared to this, \textit{GIM-RL-Fusion} aided by $\boldsymbol{s}'_k$  is trained to extract ARs in the first $200$ episodes, and continues to stably extract the large numbers of ARs.

\subsection{Results for Agent Transfer}
\label{sec:exp_transfer}

We investigate agent transfer where an agent trained on a source dataset is transferred into another agent for a target dataset which is related to the source one. For this purpose, each of the datasets in Table~\ref{tbl:datasets_stats} is split into two parts, especially, the first $60$\% of transactions and the remaining $40$\% constitute the source and target partitions, respectively. As shown in Table~\ref{tbl:source_target}, one characteristic is that the number of distinct items in the source partition ($M_{src}$) is different from the one in the target partition ($M_{tgt}$). That is, some items are included only in the source or target partition. We perform and test agent transfer between these source and target partitions.

\begin{table}[htbp]
	\caption{The difference between the number of distinct items in the source partition ($M_{src}$) and the one in the target partition ($M_{tgt}$) for each of the experimental datasets.}
	\centering
	\begin{tabular}{|r||c|c|c|c|c|} \hline
		& Chess     & Mushroom & Accidents\_$10$ & Connect   & Pumsb \\ \hline \hline
		$M_{src}$    & $72$     & $78$           & $325$  & $127$ & $1981$ \\ \hline
		$M_{tgt}$    & $75$      & $106$        & $308$  & $129$  & $1951$ \\ \hline
	\end{tabular}
	\label{tbl:source_target}
\end{table}

Let $DQN_{src}$ and $DQN_{tgt}$ be agents that are defined for the source and target partitions of a dataset, respectively. Based on the experimental results in the previous section, both of $DQN_{src}$ and $DQN_{tgt}$ are based on \textit{GIM-RL-Fusion} that stably yields high performances. In addition, as with the previous section, DQNs with the same structure are used for extracting HUIs and ARs, and simpler DQNs are for FIs. Our idea of agent transfer is very simple. All the parameters of $DQN_{src}$ trained on the source partition are transferred to $DQN_{tgt}$, except the first and output layers that need to treat the difference of distinct items between the source and target partitions. That is, $DQN_{src}$ and $DQN_{tgt}$ have the same structure with the same parameters except their first and output layers. Each unit in $DQN_{src}$'s first layer has a weight for each item. More precisely, it is used to weight $\boldsymbol{s}_k$'s value, which represents the usefulness of changing the item's inclusion in $X(\boldsymbol{b}_k)$ (the itemset defined by $\boldsymbol{b}_k$). Thus, weights of $DQN_{src}$'s first layer for items that are included in both the source and target partitions are replicated on $DQN_{tgt}$'s first layer. On the other hand, weights of $DQN_{src}$'s first layer for items included only in the source partition, are discarded. For items included only in the target partition, new randomly initialised weights are added to $DQN_{tgt}$'s first layer. The same replication, discarding and initialisation of weights are carried out for $DQN_{src}$'s and $DQN_{tgt}$'s output layers, which individually output $\boldsymbol{q}_k$ to select an action for changing the inclusion of an item in $X(\boldsymbol{b}_k)$. After the above-mentioned agent transfer, $DQN_{tgt}$ is retrained on the target partition. Assuming that the source and target partitions have similar relations among items and parameters transferred from $DQN_{src}$ to $DQN_{tgt}$ are useful for capturing those relations, retraining $DQN_{tgt}$ is expected to be much faster than training a DQN from scratch on the target partition. For simplicity, the latter DQN is called $DQN_{scratch}$.

\begin{figure*}[htbp]
	\centering
	\includegraphics[width=\linewidth]{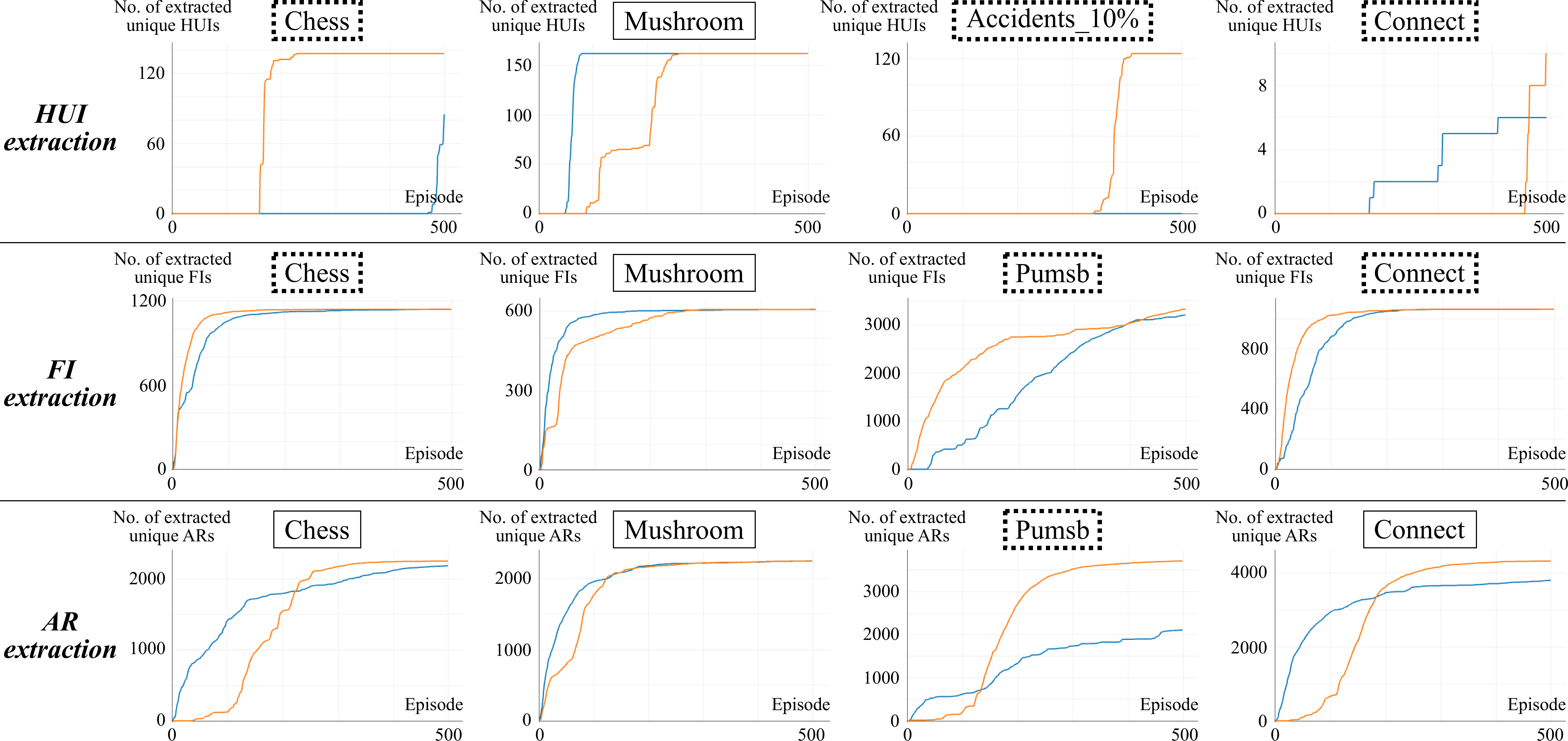}
	\caption{A comparison between the increase of the number of unique itemsets extracted by a transferred agent $DQN_{tgt}$ (orange) and the one by an ordinarily trained agent $DQN_{scratch}$ (blue) over the passage of $500$ episodes for each dataset and each itemset extraction task.}
	\label{fig:exp_transfer}
\end{figure*}

In Fig.~\ref{fig:exp_transfer}, $DQN_{tgt}$ is compared to $DQN_{scratch}$ on the target partition of each dataset. Here, the increase of the number of unique itemsets extracted by $DQN_{tgt}$ over $500$ episodes is plotted in orange, and such an increase for $DQN_{scratch}$ is drawn in blue. That is, Fig.~\ref{fig:exp_transfer} visualises how fast each of $DQN_{tgt}$ and $DQN_{scratch}$ is trained to extract itemsets in the target partition. As illustrated in this figure, in all the $12$ cases, $DQN_{tgt}$ finally extracts more itemsets or at least the same number of itemsets to $DQN_{scratch}$. Especially, the $7$ cases depicted by the dotted-line rectangles objectively indicate that $DQN_{tgt}$ extracts many itemsets of a target type more quickly than $DQN_{scratch}$. This suggests the potential effectiveness of agent transfer to realise $DQN_{tgt}$'s efficient itemset extraction with the help of $DQN_{src}$.

In addition, let us focus on Chess, Accidents\_$10$\%, Connect in the HUI extraction and Pumsb and Connect in the AR extraction in Fig.~\ref{fig:exp_transfer}. In each of these cases, $DQN_{tgt}$ extracts no or few itemsets at the beginning, but once it starts to extract itemsets, it becomes to extract many itemsets very fast. This can be thought as follows: Although parameters transferred from $DQN_{src}$ to $DQN_{tgt}$ are useful, they are not directly compatible with a target partition. But, once these parameters are adapted to the target partition, their usefulness brings in extracting many itemsets very quickly. To improve agent transfer in terms of this parameter adaptation, we plan to explore model-based RL that uses an internal model summarising an environment (i.e., dataset) and adaptively updates this model depending on changes of the environment~\cite{D_Ha}.

\subsection{Runtime Analysis}
\label{sec:exp_runtime}

Table~\ref{tbl:runtime} presents the runtimes of \textit{GIM-RL-Fusion} for each itemset extraction task. These runtimes are measured on a desktop PC equipped with Intel Core i9-9900K (3.60GHz), $32$GB RAM and NVIDIA GeForce RTX 2080Ti. \textit{GIM-RL-Fusion} implemented with Pytorch is run on Ubuntu 20.04-LTS. All the source codes for \textit{GIM-RL-Fusion} as well as the other agents are available on our Github repository, as described in Appendix~\ref{sec:app_hyper_param}. As can be seen from Table~\ref{tbl:runtime}, \textit{GIM-RL-Fusion} requires at least more than $30$ minutes to finish one task. But, we believe that this slowness is surpassed by \textit{GIM-RL-Funsion}'s great flexibility that any type of itemsets can be extracted as long as a reward for the type can be defined. In other words, \textit{GIM-RL-Funsion} can extract any user-defined type of itemsets with no need to develop a specialised data structure or algorithm for the type.

\begin{table}[htbp]
	\caption{\textit{GIM-RL-Fusion}'s runtimes expressed in the form of ``hh:mm:ss''. Column ``Acc. / Pumsb'' means that Accident\_$10$\% is used for the HUI extraction, and Pumsb is used for the FI and AR extractions.}
	\centering
	\begin{tabular}{|r||c|c|c|c|} \hline
		Type      & Chess        & Mushroom & Acc. / Pumsb  & Connect   \\ \hline \hline
		HUI       & 00:34:09     & 00:31:42      & 04:40:53           & 12:45:10  \\ \hline
		FI          & 00:31:56     & 00:33:34      & 00:50:06           & 00:37:29    \\ \hline
		AR        & 00:50:20     & 00:50:10       & 01:07:26          & 00:55:21     \\ \hline
	\end{tabular}
	\label{tbl:runtime}
\end{table}

The main reason for \textit{GIM-RL-Fusion}'s slow runtime is the need of scanning a dataset to compute a state $\boldsymbol{s}_k$ and a reward $r_k$, as illustrated by the non-filled arrows in Fig.~\ref{fig:overview}. Here, $\boldsymbol{s}_k$ and $r_k$ are computed at each step and \textit{GIM-RL-Fusion} (or more generally GIM-RL) involves $E$ episodes each of which consists of $K$ steps, as seen from Algorithm~\ref{alg:gim-rl} in Appendix~\ref{sec:app_pseudo}. Thus, $\boldsymbol{s}_k$ and $r_k$ are computed in total $EK$ times. Furthermore, $M$ values in $\boldsymbol{s}_k$ are based on  interestingness measure values for different itemsets, each of which is created by changing one item's inclusion in the itemset $X(\boldsymbol{b}_k)$ defined by the current bit-vector $\boldsymbol{b}_k$. Since the computation of an interestingness measure value requires one scan of the dataset, the computation of $\boldsymbol{s}_k$ involves $M$ scans. Moreover, $r_k$ needs the interestingness measure value for the updated itemset $X(\boldsymbol{b}_{k+1})$, so its computation also involves one dataset scan. Hence, the computation of $\boldsymbol{s}_k$ and $r_k$ at each step demands $M+1$ dataset scans. Letting $scan\_time$ be an approximate time required for one scan of the dataset, the computational complexity of \textit{GIM-RL-Fusion} can be expressed as $O\left( EK \left(M+1\right) \times scan\_time \right)$. Compared to $scan\_time$, the other processes shown in Algorithm~\ref{alg:gim-rl} of Appendix~\ref{sec:app_pseudo} require negligible times.

The reduction of $scan\_time$ is obviously crucial for speeding up \textit{GIM-RL-Fusion}. But, we will not develop a specialised data structure or algorithm to accomplish fast scan of a dataset. Instead, inspired by the work in \cite{I_Bello}, we will explore to encode the dataset into a neural network. Its input is an itemset given as a query, and the output is an approximate interestingness measure value for the itemset. This kind of speed-up of dataset scan is especially essential to extend \textit{GIM-RL-Fusion} for more complex patterns like sequential or trajectory patterns. While an itemset can be defined by a bit-vector, the representation of a sequential pattern needs a ``bit-matrix'' where one row is a bit-vector representing an itemset observed at one time point, and  the one of a trajectory pattern requires a ``bit-tensor'' where each bit-matrix at one time point indicates the x-y location of an object. The above-mentioned dataset encoding is necessary for fast querying a dataset in terms of these sequential or trajectory patterns.

\section{Conclusion and Future Work}
\label{sec:conc}

In this paper, we introduced GIM-RL that offers a unified RL framework to extract various types of itemsets only by changing a reward definition. The general effectiveness of GIM-RL is verified through the experiments on the HUI, FI and AR extractions. The experimental results also suggest one remarkable potential of GIM-RL, namely agent transfer, which realises efficient itemset mining on a dataset with the help of an agent trained on another related dataset.

\begin{figure}[htbp]
	\centering
	\includegraphics[width=\linewidth]{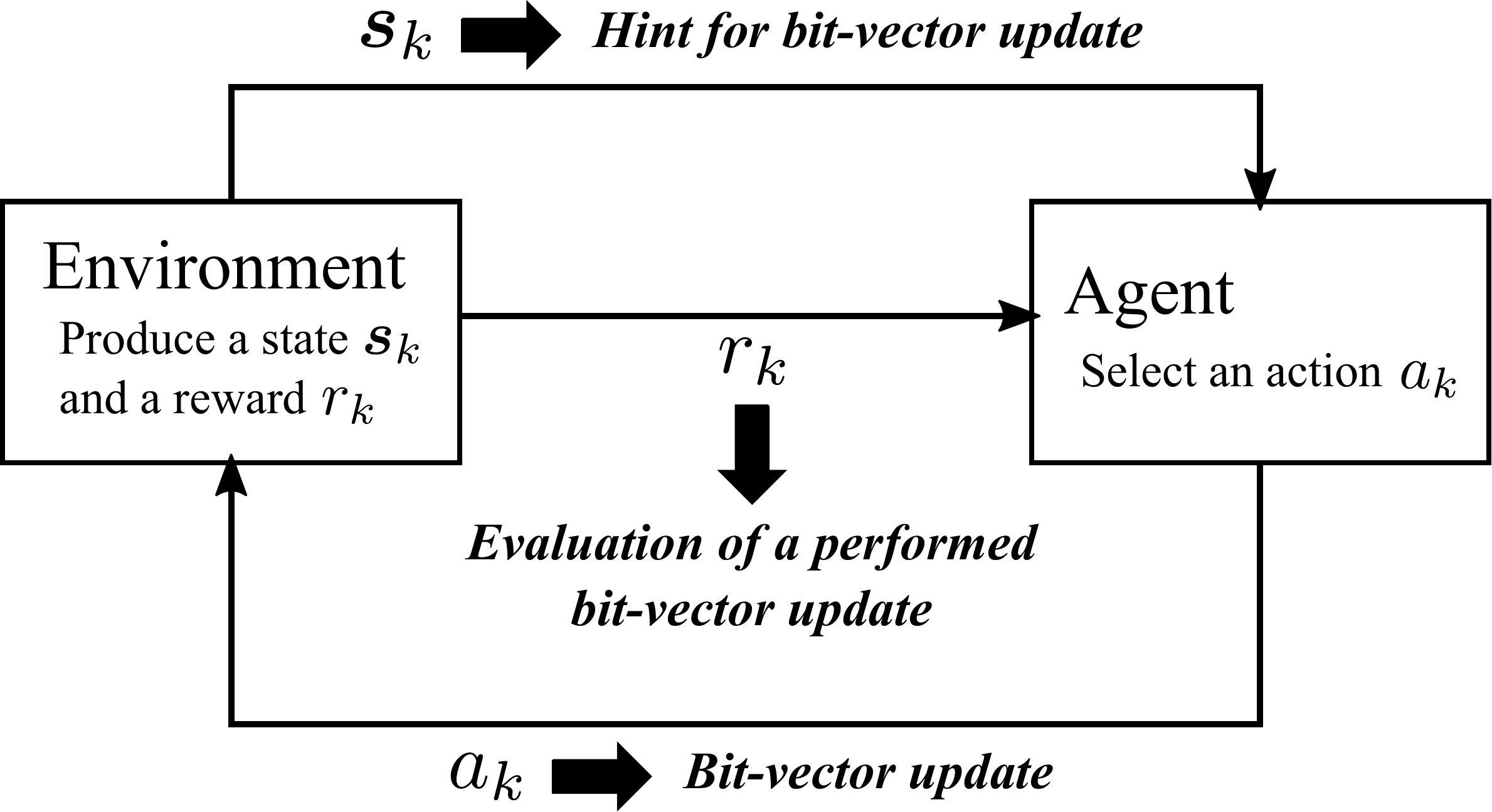}
	\caption{A conceptual illustration of how GIM-RL is formulated in the framework of RL.}
	\label{fig:gim_rl_on_rl}
\end{figure}

Before moving to detailed descriptions of our future work, let us clean up the main concept of how GIM-RL is formulated in the framework of RL by referring to Fig.~\ref{fig:gim_rl_on_rl}. GIM-RL presented in this paper is based on the most basic RL where an environment produces a state $\boldsymbol{s}_k$ based on which an agent takes an action $a_k$. Then, from the environment, the agent receives a reward $r_k$ as an evaluation of taking $a_k$ at $\boldsymbol{s}_k$ as well as the updated state $\boldsymbol{s}_{k+1}$ to take the next action $a_{k+1}$. Under this iterative interaction, RL aims to train the agent so that it can take a sequence of actions to maximise cumulative rewards. GIM-RL is formulated by defining the following $\boldsymbol{s}_k$, $a_k$ and $r_k$, as highlighted in bold italic font in Fig.~\ref{fig:gim_rl_on_rl}. First, an itemset is defined by a bit-vector indicating the inclusion of each of distinct items. Based on this, $\boldsymbol{s}_k$ is defined as a hint for how to update the bit-vector to form an itemset of a target type. In addition, $a_k$ is defined as an update of the bit-vector based on $\boldsymbol{s}_k$, and $r_k$ is an evaluation of whether the updated bit-vector expresses an itemset of the target type. According to the above-mentioned formulation, GIM-RL trains an agent that maximises cumulative rewards, namely extracts as many itemsets of the target type as possible.

Since our main purpose in this paper is to empirically verify the possibility and potential of GIM-RL, its current implementation is simple and involves many research topics that need further investigations in the future. Below, some of them are outlined in association with the components of GIM-RL's formulation in Fig.~\ref{fig:gim_rl_on_rl}. As an extension of an environment, dataset encoding described in Section~\ref{sec:exp_runtime} aims to speed-up the computation of $\boldsymbol{s}_k$ and $r_k$. Besides, model-based RL in Section~\ref{sec:exp_transfer} focuses on building a model that simulates the dynamics of an environment to generate ``virtual'' experiences, which are useful for faster agent training. Also, an agent can be enhanced by adopting a planning mechanism to infer itemsets that will be possibly obtained at the multiple-step-ahead future~\cite{R_Pascanu}. Moreover, an agent can be extended to take continuous-valued actions to extract itemsets that consist of items representing quantitative features (e.g., age, height and weight).

Apart from an environment and agent, it is also an important topic to design a new or more sophisticated state, action and reward. Especially, one invaluable attempt is to design a reward for extracting a previously unexplored type of itemsets. Another key topic is the adoption or development of a more advanced RL algorithm compared to Q-learning used in this paper (although the components in Fig.~\ref{fig:gim_rl_on_rl} are not explicitly associated with this topic). For example, training more accurate $Q(\boldsymbol{s}, a; \varTheta)$ is possible using double Q-learning where the selection and evaluation of an action are done by separate neural networks~\cite{H_Hasselt}. In addition, a policy gradient method that directly learns an action selection policy without relying on $Q(\boldsymbol{s}, a; \varTheta)$ is useful for training an agent that takes continuous-valued actions~\cite{V_Lavent}. Finally, we believe that GIM-RL opens a new research direction towards so-called ``learning-based itemset mining'' which involves many interesting topics described above.

\appendices

\newpage


\section{Pseudo-code of GIM-RL}
\label{sec:app_pseudo}

Algorithm~\ref{alg:gim-rl} shows a pseudo-code of GIM-RL to train an agent $Q(\boldsymbol{s}, a; \varTheta)$ that extracts itemsets satisfying $\varphi (X) \geq \xi$ from a dataset $\mathcal{D}$. Overall, as expressed by the double for-loop in lines $6$-$22$, the agent is trained through $E$ episodes consisting of $K$ steps. As seen at lines $10$-$12$, at each step, the agent receives a state $\boldsymbol{s}_k$ and computes $\boldsymbol{q}_k$ to decide an action $a_k$ for updating the bit-vector $\boldsymbol{b}_k$ into $\boldsymbol{b}_{k+1}$. After checking the quality of the itemset $X(\boldsymbol{b}_{k+1})$ defined by $\boldsymbol{b}_{k+1}$ at lines $13$-$15$, the environment generates a reward $r_k$ and a new state $\boldsymbol{s}_{k+1}$ at lines $16$ and $17$.  The tuple $\left(\boldsymbol{s}_k, a_k, r_k, \boldsymbol{s}_{k+1}\right)$ is then stored into a replay memory $\mathcal{P}$ as one experience of the agent at line $18$. Subsequently, the agent's parameters $\varTheta$ are updated at line $19$, while as written at line $21$ the target network's parameters $\varTheta^{-}$ are updated every $e^{-}$ episodes by copying $\varTheta$ to $\varTheta^{-}$. For \textit{GIM-RL-Fusion} in Section~\ref{sec:exp}, $\boldsymbol{q}_k$ at lines $10$ and $11$ is replaced with $\boldsymbol{q}'_k = \lambda \boldsymbol{s}'_k + (1 - \lambda) \boldsymbol{q}_k$.

\begin{algorithm}[htbp]
	\caption{GIM-RL (Generic Itemset Mining based on Reinforcement Learning)}
	\label{alg:gim-rl}
	\begin{flushleft}
		\textbf{Input} Dataset $\mathcal{D}$, interestingness measure $\varphi (X)$, threshold $\xi$\\
		\textbf{Output} A set $\mathcal{X}$ containing itemsets meeting $\varphi (X) \geq \xi$
	\end{flushleft}
	\begin{algorithmic}[1]
		\STATE Initialise $Q(\boldsymbol{s}, a; \varTheta)$ with He's parameter initialisation~\cite{K_He}
		\STATE Initialise a target network as $Q(\boldsymbol{s}, a; \varTheta^{-}) = Q(\boldsymbol{s}, a; \varTheta)$
		\STATE Initialise a replay memory as $\mathcal{P} \leftarrow \{\}$
		\STATE Filter out items that individually have no possibility to be an element of itemsets of the target type. 
		\STATE $\mathcal{X} \leftarrow \{\}$
		\FOR{$e= 1, \cdots E$}
		\STATE Randomly initialise the bit-vector $\boldsymbol{b}_1$
		\STATE Generate the first state $\boldsymbol{s}_1$ based on $\boldsymbol{b}_1$ and $\mathcal{D}$
		\FOR{$k=1, \cdots K$}
		\STATE Compute $\boldsymbol{q}_k$ by feeding $\boldsymbol{s}_k$ into $Q(\boldsymbol{s}, a; \varTheta)$
		\STATE Decide an action $a_{k}$ by the $\epsilon$-greedy strategy on $\boldsymbol{q}_k$
		\STATE Update $\boldsymbol{b}_{k}$ into $\boldsymbol{b}_{k+1}$ by $a_{k}$
		\IF{$\varphi (X(\boldsymbol{b}_{k+1})) \geq \xi$}
		\STATE $\mathcal{X} \leftarrow \mathcal{X} \ \cup \ \{ X(\boldsymbol{b}_{k+1}) \}$ \; // An itemset is extracted
		\ENDIF
		\STATE Compute a reward $r_{k}$ based on $\boldsymbol{b}_{k+1}$ and $\mathcal{D}$ (Section~\ref{sec:reward})
		\STATE Compute $\boldsymbol{s}_{k+1}$ based on $\boldsymbol{b}_{k+1}$ and $\mathcal{D}$
		\STATE $\mathcal{P} \leftarrow \mathcal{P} \ \cup \{ (\boldsymbol{s}_k, a_k, r_k, \boldsymbol{s}_{k+1}) \}$
		\STATE Update $\varTheta$ of $Q(\boldsymbol{s}, a; \varTheta)$ using experiences randomly sampled from $\mathcal{P}$
		\ENDFOR
		\STATE Update $\varTheta^{-}$ of $Q(\boldsymbol{s}, a; \varTheta^{-})$ as $\varTheta^{-} = \varTheta$ (every $e^{-}$ episodes)
		\ENDFOR
		\RETURN $\mathcal{X}$
	\end{algorithmic}
\end{algorithm}

We mention the following three implementation details: First, the search space reduction before extracting itemsets is done at line $4$. Specifically, for the extraction of HUIs, items whose upper bound utilities (transaction weighted utilisations~\cite{V_Tseng,C_Ahmed,M_Liu,W_Song}) are less than the threshold $\xi$ are discarded, and for the extractions of FIs and ARs, items whose supports are less than the minimum support threshold ($\xi$ for FIs, and $\xi_1$ for ARs) are eliminated. Second, at line $7$, $\boldsymbol{b}_1$ in each episode is obtained by the random bit-vector initialisation described in Section~\ref{sec:framework}. For this initialisation, we assume that the more frequently an item occurs in $\mathcal{D}$, the more likely it is to be included in HUIs, FIs or ARs. The above-mentioned search space reduction and random bit-vector initialisation need to be modified when targeting another type of itemsets like infrequent itemsets. Last, to help the agent explore a variety of itemsets, line $11$ shows that $a_k$ is decided using the $\epsilon$-greedy strategy. Here, $a_k$ is chosen as the action corresponding to the highest value in $\boldsymbol{q}_k$ with the probability $1 - \epsilon$, while $a_k$ is set to the action to change the value of a randomly selected dimension in $\boldsymbol{b}_k$ with the probability $\epsilon$.

\section{DQN Architectures}
\label{sec:app_dqn_arch}

\begin{figure}[htbp]
	\centering
	\includegraphics[width=\linewidth]{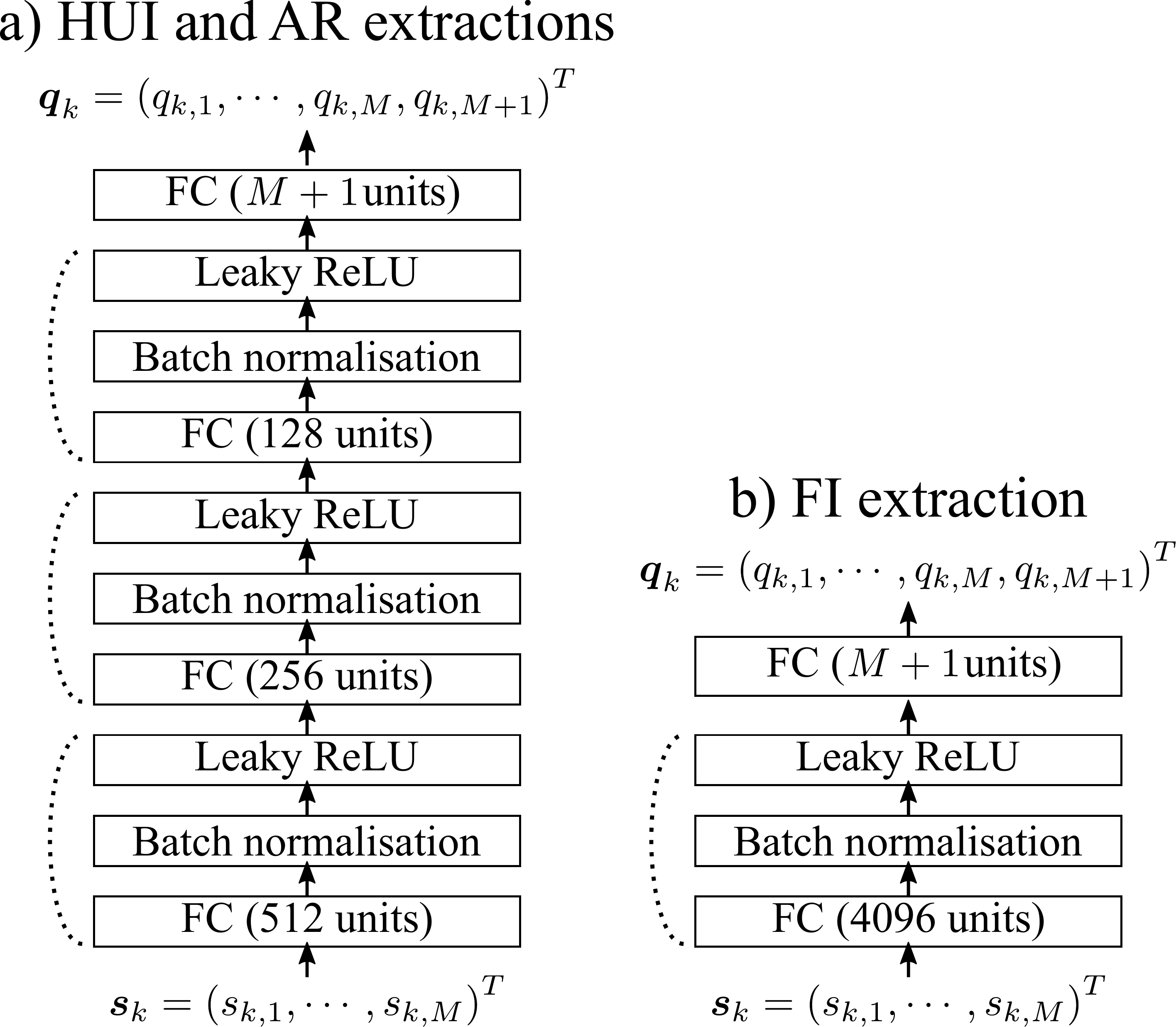}
	\caption{DQN architectures used for the HUI, FI and AR extractions.}
	\label{fig:dqn_arch}
\end{figure}

Figs.~\ref{fig:dqn_arch} (a) and (b) show the DQN architecture for extracting HUIs and ARs and the one for FIs, respectively. These architectures are common with respect to the input and output layers. The input layer accepts an $M$-dimensional state vector $\boldsymbol{s}_k = \left(s_{k,1}, \cdots, s_{k,M}\right)^T$ where $s_{k,m}$ ($1 \leq m \leq M$) represents an estimated usefulness of changing the $m$th item's inclusion in the itemset $X(\boldsymbol{b}_k)$. The output layer produces an $(M+1)$-dimensional vector $\boldsymbol{q}_k = ( q_{k,1}, \cdots, q_{k,M+1})^T$ where $q_{k,m}$ ($1 \leq m \leq M$) and $q_{k,M+1}$ indicate an approximate quality of the action to change the $m$th item's inclusion and the one for the random bit-vector  initialisation, respectively. As illustrated by the three dotted arcs in Fig.~\ref{fig:dqn_arch} (a), the DQN for the HUI and AR extractions has three blocks each of which consists of a Fully-Connected (FC) layer, a batch normalisation layer and an activation layer defined by leaky ReLU (Rectified Linear Unit). Similarly, as shown in Fig.~\ref{fig:dqn_arch} (b), the DQN for the FI extraction is comprised of one block with an FC layer having a large number of units. Based on our preliminary experiments, the structural complexity of a DQN may be related to the complexity of a target type. That is, FIs are simpler than HUIs and ARs, so a simpler DQN seems enough for extracting FIs compared to the one for HUIs and ARs.

We describe details of how to compute $s_{k,m}$ in $\boldsymbol{s}_k$. First of all, $\varphi (X)$ has a huge value range. For example, assuming that $\varphi(X)$ outputs the support of an itemset $X$, one itemset may have a support of $1000$ while the support of a rare itemset may be $1$. Thus, if $s_{k,m}$ is defined directly as $\varphi(X(\boldsymbol{b}'_k))$ where $X(\boldsymbol{b}'_k)$ is the itemset created by changing the $m$th item's inclusion in $X(\boldsymbol{b}_k)$ (please see Section~\ref{sec:framework}), $s_{k,m}$'s value range can be huge, and $\boldsymbol{s}_k$ is often very biased in the sense that only some dimensions have very large values. To alleviate this bias, $s_{k,m}$ is computed by normalising $\varphi(X(\boldsymbol{b}'_k))$ as follows:
\begin{equation}
	s_{k,m} = \log \left( \frac{\varphi(X(\boldsymbol{b}'_k))}{Z} + 1 \right),
	\label{eq:state_comp}
\end{equation}
where $Z$ is a normalisation factor to put $\varphi(X(\boldsymbol{b}'_k)) / Z$ between $0$ and $1$. For this, using $Z$ is common in itemset mining, for instance, in FI extraction, $Z$ is the number of transactions in $\mathcal{D}$ to compute the ``relative'' support of an itemset~\cite{R_Agrawal,J_Han,P_Viger_ASO}, and in HUI extraction, relative utilities are calculated by setting $Z$ to the sum of utilities for all transactions (i.e., sum of transaction utilities) in $\mathcal{D}$~\cite{M_Liu,V_Tseng,W_Song,W_Song2}. These definitions of $Z$ are also used in Eq.~\ref{eq:state_comp}, which then takes $\log$ of $(\varphi(X(\boldsymbol{b}'_k)) / Z + 1)$ to further reduce value differences among $s_{k,1} \cdots , s_{k,M}$ in $\boldsymbol{s}_k$.

One natural extension of $\boldsymbol{s}_k$ is performed for AR extraction involving two interestingness measures $\varphi_1(X)$ and $\varphi_2(X)$ that output the support and confidence of $X$, respectively. Specifically, we compute a $2M$-dimensional state vector $\boldsymbol{s}_k$. Here, $s_{k,m}$ in the first $M$ dimensions is computed based on $\varphi_1 (X(\boldsymbol{b}'_k))$ and $Z$ being the number of transactions in $\mathcal{D}$. On the other hand, $s_{k,m}$ in the last $M$ dimensions is computed based on $\varphi_2 (X(\boldsymbol{b}'_k))$ and $Z=1$, because $\varphi_2 (X(\boldsymbol{b}'_k))$ representing the confidence of the rule defined by $X(\boldsymbol{b}_k)$ and $X(\boldsymbol{b}'_k)$ already lies between $0$ and $1$. Finally, $\boldsymbol{s}_k$ is fed into a batch normalisation layer to normalise values by $\varphi_1 (X(\boldsymbol{b}'_k))$ and $\varphi_2 (X(\boldsymbol{b}'_k))$, and the resulting normalised $\boldsymbol{s}_{k}$ is used as the input of the DQN architecture in Fig.~\ref{fig:dqn_arch} (a). For \textit{GIM-RL-Fusion}, the first $M$ dimensions based on $\varphi_1 (X(\boldsymbol{b}'_k))$ are used to create $\boldsymbol{s}'_k$.

\section{Hyper-parameter Setting}
\label{sec:app_hyper_param}

Referring to Algorithm~\ref{alg:gim-rl} in Appendix~\ref{sec:app_pseudo}, the number of episodes $E$ is set to $500$ for the HUI extraction and $1000$ for the FI and AR extractions in Section~\ref{sec:exp_im}, and to $500$ for all the experiments of agent transfer in Section~\ref{sec:exp_transfer}. The number of steps in one episode is always $K = 500$. The size of a replay memory $\mathcal{P}$ is set to $10000$. Each update of an agent's parameters $\varTheta$ is based on Eq.~\ref{eq:q_loss} where the discount factor $\gamma$ is set to $0.95$. This update of $\varTheta$ is done using RAdam~\cite{L_Liu} (with the initial learning rate $\alpha = 0.001$) as an optimiser on a mini-batch of $512$ experiences randomly sampled from $\mathcal{P}$. Furthermore, $e^{-}$ which is the episode period to update a target network's parameters $\varTheta^{-}$ is set to $5$, that is, $\varTheta^{-}$ is updated every $e^{-}=5$ episodes.

$\lambda$ in $\boldsymbol{q}'_k = \lambda \boldsymbol{s}'_k + (1 - \lambda) \boldsymbol{q}_k$ of \textit{GIM-RL-Fusion} is dynamically changed. As described in Section~\ref{sec:exp_im}, at the beginning of training an agent, $\lambda$ is large in order to facilitate finding positive itemsets based on $\boldsymbol{s}'_k$, and is gradually reduced to prioritise the agent's exploitation based on the trained $Q (\boldsymbol{s}, a; \varTheta)$. Inspired by the $\epsilon$-greedy strategy, we implement this dynamic change of $\lambda$ as follows:
\begin{equation}
	\lambda = \lambda_{end} + (\lambda_{start} - \lambda_{end}) \exp \left( -\frac{k_{total}}{\varDelta} \right),
	\label{eq:lambda}
\end{equation}
where $\lambda_{start}$ and $\lambda_{end}$ are the maximum and minimum values of $\lambda$, respectively. $k_{total}$ is the total number of steps the agent has experienced, that is, $k_{total} = e \times K + k$ for the $k$th step in the $e$th episode. In Eq.~\ref{eq:lambda}, $\lambda$ is $\lambda_{start}$ when $k_{total} = 0$, and gradually converges to $\lambda_{end}$ as $k_{total}$ increases. $\varDelta$ is a hyper-parameter to control the speed of this convergence.

\begin{figure*}[htbp]
	\centering
	\includegraphics[width=\linewidth]{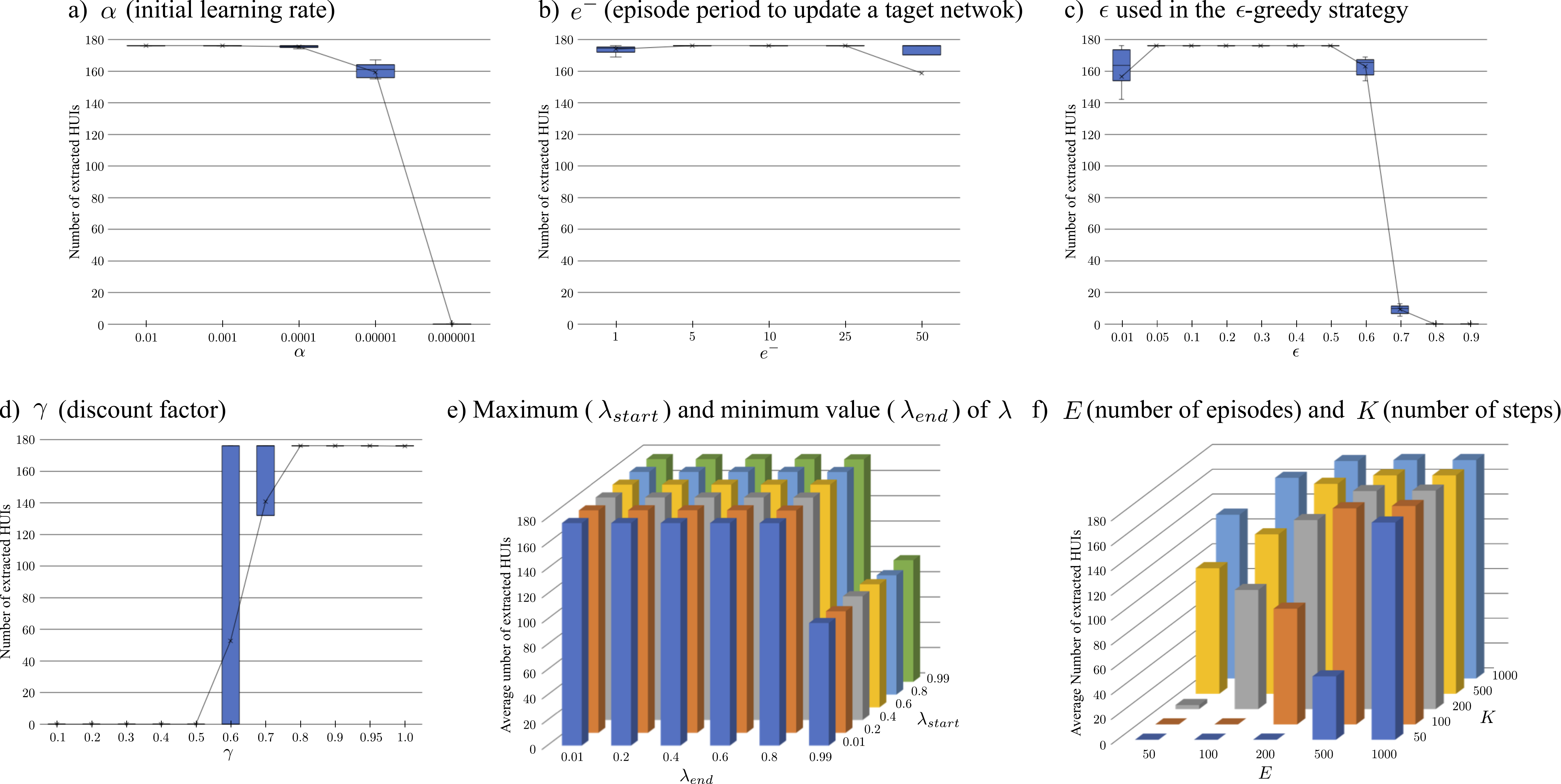}
	\caption{Transitions of numbers of HUIs extracted from Chess by changing one or two hyper-parameters of \textit{GIM-RL-Fusion}.}
	\label{fig:hyper_param_effects}
\end{figure*}

First, $\varDelta$ is set to 200 in all the experiments in Section~\ref{sec:exp}. In Section~\ref{sec:exp_im}, $\lambda_{start}=0.999$ and $\lambda_{end}=0.5$ are used for extracting HUIs and ARs, and $\lambda_{start}=0.999$ and $\lambda_{end}=0.6$ are for FIs. For these settings of $\lambda_{start}$ and $\lambda_{end}$, we think that $\boldsymbol{s}_k$ provides ``dataset-dependent'' information for changing each item's inclusion in $X(\boldsymbol{b}_k)$ by actually checking transactions in a dataset, while $\boldsymbol{q}_k$ represents more general information obtained from a DQN. According to this thought, HUIs and ARs are more complex than FIs, so the smaller $\lambda_{end} = 0.5$ is used to put a higher priority on $\boldsymbol{q}_k$.

For the experiments of agent transfer in Section~\ref{sec:exp_transfer}, $\lambda_{start}$ and $\lambda_{end}$ are set depending on source and target datasets, and itemset types. For the HUI extraction, $\lambda_{start} = 0.999$ and $\lambda_{end} = 0.5$ are used to train an agent on a source dataset, and $\lambda_{start} = \lambda_{end} = 0.5$ (i.e., $\lambda$ is constant at $0.5$) is used on a target dataset in order to take more advantage of $\boldsymbol{q}_k$ obtained from the source dataest. For the FI extraction, $\lambda_{start} = 0.999$ and $\lambda_{end} = 0.6$ are used for both source and target datasets, to take more account of dataset-dependent information for the target dataset. For the AR extraction, we use $\lambda_{start} = \lambda_{end} = 0.5$ on a source dataset to obtain general information about items as $\boldsymbol{q}_k$, and then $\lambda_{start} = 0.999$ and $\lambda_{end} = 0.5$ are used to gradually adapt $\boldsymbol{q}_k$ to a target dataset by considering dataset-dependent information. We will explore a more sophisticated approach to define $\lambda_{start}$ and $\lambda_{end}$ based on statistical characteristics of source and target datasets.

Finally, all the source codes (including the codes for data download) used in this paper are available on our Github repository\footnote{\url{https://github.com/Wisteria30/GIM-RL}}.

\section{Effects of Hyper-parameters}
\label{sec:app_hyper_param_effects}

This section is devoted to examining effects of GIM-RL's hyper-parameters on itemset mining results. Due to the computational cost and the presentation brevity, the following experiments especially address the HUI extraction on Chess, where the maximum number of extractable HUIs is $176$ based on the threshold $\xi$ in Table~\ref{tbl:hui_result} of Section~\ref{sec:exp_im}. In addition, \textit{GIM-RL-Fusion} is particularly used because it stably achieved the top or nearly top performance on all the experiments.

\begin{figure*}[htbp]
	\centering
	\includegraphics[width=\linewidth]{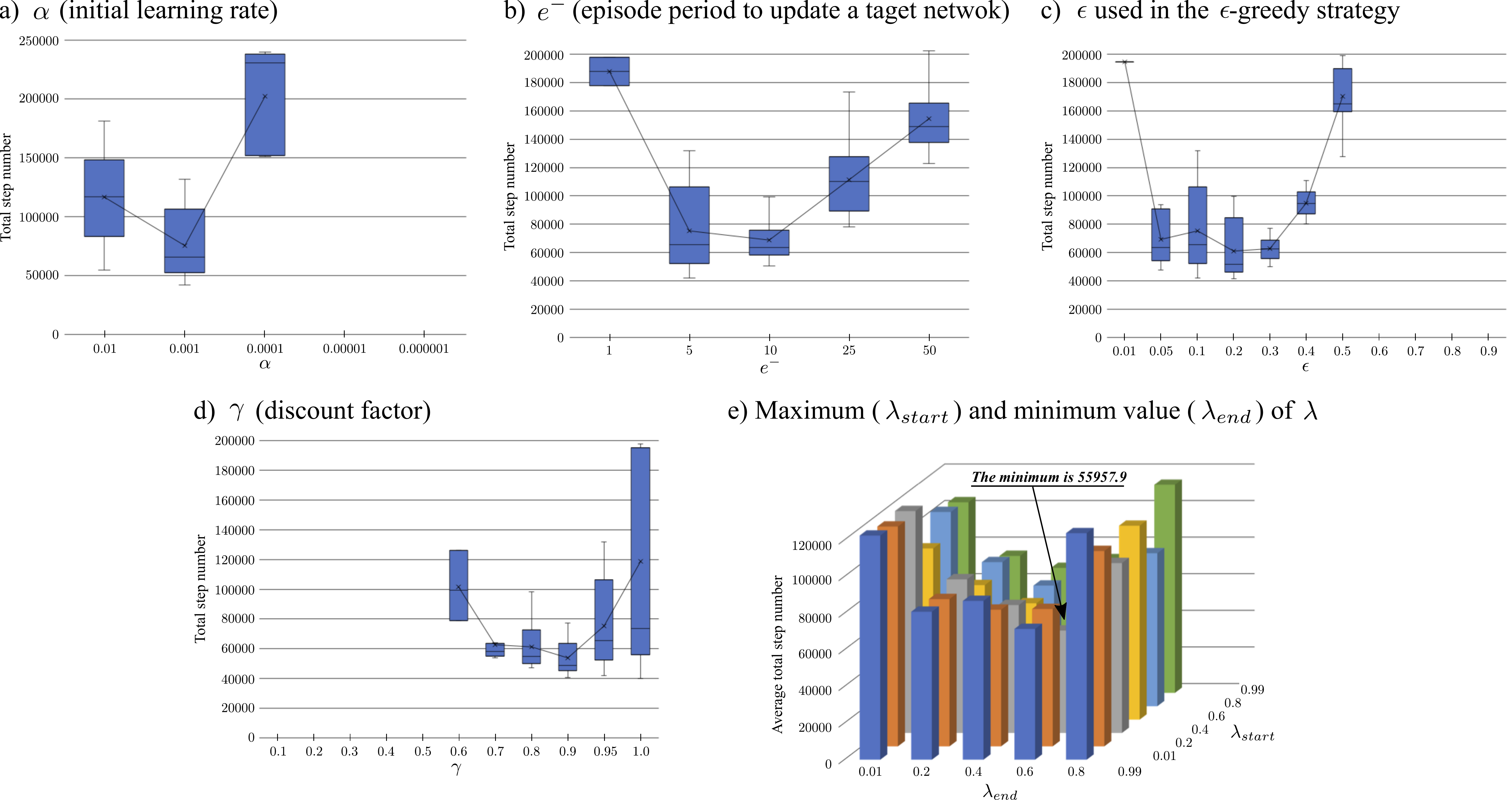}
	\caption{Transitions of total step numbers required to extract all the HUIs in Chess by changing one or two hyper-parameters of \textit{GIM-RL-Fusion}.}
	\label{fig:hyper_param_effects_steps}
\end{figure*}

Fig.~\ref{fig:hyper_param_effects} illustrates transitions of numbers of HUIs extracted by changing one or two hyper-parameters of \textit{GIM-RL-Fusion}. Note that all hyper-parameters except the changed ones are kept the same to the ones in Section~\ref{sec:exp_im} (please see the previous section for the concrete values of the unchanged hyper-parameters). Each of Figs.~\ref{fig:hyper_param_effects} (a) to (d) presents a transition of numbers of HUIs extracted using hyper-parameter values shown on the horizontal axis. Here, considering the randomness involved in \textit{GIM-RL-Fusion}, a box-plot is adopted by running it $10$ times for each hyper-parameter value. The line is drawn by connecting the average numbers of HUIs extracted in $10$ runs for neighbouring hyper-parameter values\footnote{The line lying outside the box for $e^{-} = 50$ in Fig.~\ref{fig:hyper_param_effects} (b) is caused by one exceptionally small number of extracted HUIs. In addition, the long box for $\gamma=0.6$ in Fig.~\ref{fig:hyper_param_effects} (d) results from the mixture of runs where all the HUIs are extracted and the ones where no HUI is extracted.}. Figs.~\ref{fig:hyper_param_effects} (e) and (f) employ three-dimensional histograms that are individually based on hyper-parameter pairs defined by the bottom plane. The vertical axis of each histogram indicates the average of numbers of HUIs extracted in $10$ runs with each hyper-parameter pair (the standard deviation is omitted for the presentation brevity).

Let us focus on Figs.~\ref{fig:hyper_param_effects} (b) regarding $e^{-}$ to handle the episode period for updating a target network, (c) regarding $\epsilon$ to control the probability of a random action in the $\epsilon$-greedy strategy,  and (e) regarding $\lambda_{start}$ and $\lambda_{end}$ to adjust the mixing weight between an extended state $\boldsymbol{s}_k'$ and a vector of estimated action qualities $\boldsymbol{q}_k$. For these hyper-parameters, all the $176$ HUIs are extracted as long as they are set to moderate values like $5$-$25$ for $e^{-}$, $0.05$-$0.5$ for $\epsilon$ and $0.01$-$0.8$ for $\lambda_{start}$ and $\lambda_{end}$. The initial learning rate $\alpha$ in Fig.~\ref{fig:hyper_param_effects} (a) needs special attention like usual training of a neural network. Fig.~\ref{fig:hyper_param_effects} (a) indicates that setting $\alpha$ to a very small value slows down agent training, consequently a much smaller number of HUIs can be extracted in the same episode and step numbers. Fig.~\ref{fig:hyper_param_effects} (d) treats a discount factor $\gamma$ to control the degree of lowering rewards at future steps. More intuitively, $\gamma$ manages the future range that an agent considers to select an action at the current step. Fig.~\ref{fig:hyper_param_effects} (d) exhibits that more HUIs are extracted as the further future is considered using larger $\gamma$. For Fig.~\ref{fig:hyper_param_effects} (f) concerning $E$ and $K$ to control the episode and step numbers, it is clear that the number of HUIs decreases as these hyper-parameters are set to smaller values. On the other hand, setting them to larger values leads to extracting all the HUIs although it lengthens the runtime of \textit{GIM-RL-Fusion}.

In Fig.~\ref{fig:hyper_param_effects}, all the HUIs are extracted by multiple values of a hyper-parameter. Thus, more detailed analysis is conducted by adopting another evaluation criteria about the efficiency of itemset mining. Based on Section~\ref{sec:exp_runtime}, when targeting a particular dataset, the time needed for each step of GIM-RL is assumed to be nearly the same, because the number of distinct items $M$ is fixed and the time $scan\_time$ required for one dataset scan should be approximately constant. Hence, we define the efficiency of GIM-RL's itemset mining as the total number of steps required to extract a certain number of itemsets, especially, all the itemsets. For simplicity, this number is abbreviated as a \textit{total step number}. To be precise, it is computed as $e' \times K + k'$ if all the itemsets are extracted at the $k'$th step in $e'$th episode.

Fig.~\ref{fig:hyper_param_effects_steps} shows transitions of total step numbers required to extract all the HUIs in Chess by changing each of $\alpha$, $e^{-}$, $\epsilon$, $\gamma$ and the pair of $\lambda_{start}$ and $\lambda_{end}$ (a graph for $E$ and $K$ is not created because they obviously have no influence on total step numbers). Since a total step number can be calculated only in the case where all the HUIs are extracted, Fig.~\ref{fig:hyper_param_effects_steps} focuses on hyper-parameter values leading to this case in Fig.~\ref{fig:hyper_param_effects}. That is, nothing is depicted in Fig.~\ref{fig:hyper_param_effects_steps} for hyper-parameter values with which all the HUIs are not extracted in Fig.~\ref{fig:hyper_param_effects}. In each of Figs.~\ref{fig:hyper_param_effects_steps} (a) to (d), \textit{GIM-RL-Fusion} is run $10$ times and a box-plot of total step numbers is drawn based on the runs where all the HUIs are extracted. The three-dimensional histogram in Fig.~\ref{fig:hyper_param_effects_steps} (e) displays the average of total step numbers obtained in $10$ runs using each pair of $\lambda_{start}$ and $\lambda_{end}$.

Fig.~\ref{fig:hyper_param_effects_steps} shows that even if multiple hyper-parameter values result in extracting all the HUIs, total step numbers for them are quite different. That is, GIM-RL's runtime to extract all the itemsets significantly varies depending on hyper-parameter values. And, analysis of total step numbers is useful for finding hyper-parameter values with which many itemsets can be efficiently extracted in a short runtime. With respect to this, it is often infeasible or takes very long time to extract all the itemsets in a huge real-world dataset, so computing total step numbers for different hyper-parameter values is impossible or impractical. Nevertheless, random sampling of transactions in the original dataset can be performed to create a moderate-size dataset for which total step numbers are computed reasonably. This way, even for real-world datasets, it is possible to tune hyper-parameters so as to find a good tradeoff between GIM-RL's runtime (i.e., total step number) and the number of extracted itemsets.

\section{Small Remark about Mushroom}
\label{sec:app_mushroom}

The Web page\footnote{\url{http://www.philippe-fournier-viger.com/spmf/index.php?link=datasets.php}} of SPMF library~\cite{P_Viger_SPMF} provides two versions of Mushroom, one for the HUI extraction and the other for the FI and AR extractions. Regarding the number of distinct items in a source partition $M_{src}$ and the one in a target partition $M_{tgt}$ in Table~\ref{tbl:source_target} of Section~\ref{sec:exp_transfer}, these two versions have a small difference. Mushroom for the HUI extraction is divided into the source parition with $M_{src} = 98$ and the target one with $M_{tgt} = 109$, and the source and target partitions of Mushroom for the FI and AR extractions are characterised by $M_{src} = 78$ and $M_{tgt} = 106$, respectively. Only the latter $M_{src}$ and $M_{tgt}$ are shown in Table~\ref{tbl:source_target} for ease of understanding.

\section{Abbreviations and Symbols}
\label{sec:app_abb_sym}

Below, a list of abbreviations (Table~\ref{tbl:abb_list}) and two lists of mathematical symbols (Tables~\ref{tbl:symbol_list} and \ref{tbl:symbol_list_details}) are provided to make this paper easier to understand. Especially, Tables~\ref{tbl:symbol_list} and \ref{tbl:symbol_list_details} are created by categorising symbols based on whether they are defined to explain the framework of GIM-RL in Section~\ref{sec:gim-rl} or describe details of our implementation and experiments in and after Section~\ref{sec:exp}.

\begin{table}[htbp]
	\centering
	\caption{A list of abbreviations.}
	\label{tbl:abb_list}
	\begin{tabular}{|c|l|} \hline
	Abbreviation & Expansion \\ \hline \hline
	FI  & Frequent Itemset \\ \hline
	AR & Association Rule \\ \hline
	HUI & High Utility Itemset \\ \hline
	FP-tree & Frequent Patten tree \\ \hline
	FP-growth & Frequent Pattern growth\\ \hline
	RL & Reinforcement Learning \\ \hline
	\multirow{2}{*}{GIM-RL} & Generic Itemset Mining based on Reinforcement\\
                                                    &  Learning (our proposed approach) \\ \hline
	DQN & Deep Q-Network \\ \hline     
	EC & Evolutionary Computation \\ \hline
	GA & Genetic Algorithm\\ \hline
	SPMF & Sequential Pattern Mining Framework\\ \hline
	FC layer & Fully-Connected layer \\ \hline
	leaky ReLU & leaky Rectified Linear Unit \\ \hline
	\end{tabular}
\end{table}

\begin{table}[htbp]
	\centering
	\caption{A list of mathematical symbols defined to explain the framework of GIM-RL in Section~\ref{sec:gim-rl}.}
	\label{tbl:symbol_list}
	\begin{tabular}{|c|l|}
		\multicolumn{2}{c}{Symbols related to itemset mining} \\ \hline
		Symbol & Description \\ \hline \hline
		\multirow{2}{*}{$\mathcal{D}$} & A dataset containing $N$ transactions, \\
		&  $\mathcal{D} = \{T_1, \cdots, T_N\}$   \\ \hline
		$N$      & The number of transactions in $\mathcal{D}$ \\ \hline
		$\mathcal{I}$  & The set of $M$ distinct items in $\mathcal{D}$, $\mathcal{I} = \{i_1,  \cdots, i_M\}$ \\ \hline
		$M$      & The number of items in $\mathcal{I}$ \\ \hline
		\multirow{2}{*}{$T_n$} & The $n$th transaction in $\mathcal{D}$,\\
		& $T_n  =  \{ i_{n,1}, \cdots , i_{n,|T_n|} \} \subseteq \mathcal{I}$ \\ \hline
		$i_{n,l}$            & The $l$th item in $T_n$,  $i_{n,l} \in \mathcal{I}$ \\ \hline
		$i_m$                & The $m$th item in $\mathcal{I}$ \\ \hline
		$X$                    & An itemset, $X \subseteq \mathcal{I}$ \\ \hline
		\multirow{2}{*}{$\varphi (X)$}  & $X$'s relevance to a target type (i.e., interestingness\\
		                                                          &  measure value) \\ \hline
		\multirow{2}{*}{$\xi$}   & A threshold to check the sufficiency of $X$'s relevance \\
		& ($X$ is extracted if $\varphi(X) \geq \xi$)  \\ \hline
		$p(i_{n,l})$         & The item-specific utility of $i_{n,l}$ in $T_n$ to define HUIs \\ \hline 
		$q(i_{n,l})$         & The quantity of $i_{n,l}$ in $T_n$ to define HUIs \\ \hline 
		$sup(X)$ & $X$'s support (frequency) to define FIs and ARs \\ \hline
		$X_a$      & An itemset defining the antecedent of an AR  \\ \hline 
		$X_c$      & An itemset defining the consequent of an AR  \\ \hline 
		$min\_sup$ & A minimum support threshold to define FIs and ARs\\ \hline
		$min\_conf$ & A minimum confidence threshold to define ARs\\ \hline
		\multicolumn{2}{c}{ } \\
		\multicolumn{2}{c}{ } \\		
		\multicolumn{2}{c}{Symbols related to RL for itemset mining} \\ \hline
		Symbol & Description \\ \hline \hline
		\multirow{2}{*}{$\boldsymbol{b}_k$}  & An $M$-dimensional bit-vector to define the itemset\\
		& examined at the $k$th step,  $\boldsymbol{b}_k  = \left(b_{k,1}, \cdots, b_{k,M} \right)^T$ \\ \hline
		\multirow{2}{*}{$b_{k,m}$} & The $m$th binary value in $\boldsymbol{b}_k$ representing the inclusion\\
		& of the $m$th item $i_m$, $b_{k,m} \in \{0, 1\}$ \\ \hline
		$X(\boldsymbol{b}_k)$ & The itemset defined by $\boldsymbol{b}_k$ \\ \hline
		\multirow{3}{*}{$\boldsymbol{s}_k$} & A state defined as an $M$-dimensional vector\\
		                                                                    & suggesting how to change $\boldsymbol{b}_k$ to form an itemset \\
		                                                                    & of a target type, $\boldsymbol{s}_k = \left(s_{k,1}, \cdots, s_{k,M} \right)^T$ \\ \hline
		\multirow{2}{*}{$s_{k,m}$} & The $m$th value in $\boldsymbol{s}_k$ implying the usefulness for\\
		& changing $b_{k,m}$ in $\boldsymbol{b}_k$ (i.e., $i_m$'s inclusion in $X(\boldsymbol{b}_k)$)\\ \hline
		\multirow{2}{*}{$a_k$} & An action taken by an agent at the $k$th step to update\\
		                                           & $\boldsymbol{b}_k$ into $\boldsymbol{b}_{k+1}$, $a_k \in \mathcal{A}$ \\ \hline
		\multirow{3}{*}{$\mathcal{A}$} & A set of possible actions including the change of\\
		                                                           & each item's inclusion and the random bit-vector \\
		                                                           & initialisation\\ \hline
		\multirow{2}{*}{$r_k$}                    & A reward indicating an evaluation score of taking\\
		                                                          & $a_k$ at $\boldsymbol{s}_k$\\ \hline
		\multirow{2}{*}{$Q^{*} (\boldsymbol{s}, a)$} & The optimal Q function quantifying the quality of \\
		& taking an action $a$ at a state $\boldsymbol{s}$ \\ \hline
		$K$                                  & The number of steps to update a bit-vector \\ \hline
		$\gamma$                      & A discount factor to lower future rewards\\ \hline
		$\pi$                                & A policy for action selection\\ \hline
		\multirow{2}{*}{$Q (\boldsymbol{s}, a; \varTheta)$} & A Q function approximated by a neural network \\
		                                                                                    & (i.e., DQN) defined by a set of parameters $\varTheta$ \\
		                                                                                    & ($Q (\boldsymbol{s}, a; \varTheta)$ is interchangeably called an agent) \\ \hline
		\multirow{3}{*}{$\boldsymbol{q}_k$} & An $(M+1)$-dimensional vector collecting\\
		                                                                    & $Q (\boldsymbol{s}, a; \varTheta)$s for all the $M+1$ actions
		                                                                    at the $k$th step,\\
		                                                                    & $\boldsymbol{q}_k = \left( q_{k,1}, \cdots, q_{k,M+1} \right)^T$\\ \hline
		$E$                                  & The number of episodes each consisting of $K$ steps\\ \hline
		\multirow{2}{*}{$\mathcal{P}$} & A replay memory storing an agent's recent \\
		                                                           & experiences $(\boldsymbol{s}_k,a_k,r_k,\boldsymbol{s}_{k+1})$s\\ \hline
		\multirow{2}{*}{$Q (\boldsymbol{s}, a; \varTheta^{-})$} & A target network with parameters $\varTheta^{-}$ to compute\\
                                                                                                     &  the target value $r_k + \gamma \max_{a' \in \mathcal{A}} Q(\boldsymbol{s}_{k+1}, a'; \varTheta^{-})$\\ \hline
	\end{tabular}
\end{table}

\begin{table}[htbp]
	\centering
	\caption{A list of mathematical symbols defined to describe details of our implementation and experiments in and after Section~\ref{sec:exp}.}
	\label{tbl:symbol_list_details}
	\begin{tabular}{|c|l|} \hline
		Symbol & Description \\ \hline \hline
		$e^{-}$ & The episode period to update a target network\\ \hline
		$\alpha$ & The initial learning rate in RAdam optimiser\\ \hline
		\multirow{3}{*}{$\epsilon$} & A probability that an agent takes a random action\\
		                                                   & ($\epsilon$ is used in two similar contexts, please see \textit{State-}$\epsilon$ \\
		                                                   & in Section~\ref{sec:exp_im} and GIM-RL in Appendix~\ref{sec:app_pseudo}) \\ \hline
		\multirow{2}{*}{$\boldsymbol{s}'_k$} & An extended state created by appending to $\boldsymbol{s}_k$ one\\
		                                                                     & dimension for the random bit-vector initialisation\\ \hline
		\multirow{2}{*}{$\boldsymbol{q}'_k$} & $\boldsymbol{q}_k$ combined with $\boldsymbol{s}'_k$ to advance the training of\\
		                                                                     & an agent, $\boldsymbol{q}'_k =  \lambda \boldsymbol{s}'_k + (1 - \lambda) \boldsymbol{q}_k$ \\ \hline
		\multirow{2}{*}{$\lambda$} & A dynamic weight to combine $\boldsymbol{q}_k$ with $\boldsymbol{s}'_k$ through\\
		                                                    & the passage of training an agent\\ \hline 
		$\lambda_{start}$ & The initial value of $\lambda$ corresponding to its maximum\\ \hline
		\multirow{2}{*}{$\lambda_{end}$} & The convergence value of $\lambda$ corresponding to its\\
		                                                               & minimum\\ \hline
		\multirow{2}{*}{$k_{total}$} & The total number of steps executed until the $k$th step\\ 
		                                                     &  in the $e$th episode, $k_{total} = e \times K + k$\\ \hline
		$\varDelta$ & A factor to control the speed of $\lambda$'s convergence\\ \hline
		$M_{src}$ & The number of distinct items in a source partition\\ \hline
		$M_{tgt}$ & The number of distinct items in a target partition\\ \hline
		$DQN_{src}$ & A DQN trained on a source partition\\ \hline
		\multirow{2}{*}{$DQN_{tgt}$} & A DQN trained by transferring $DQN_{src}$ to a\\
		                                                       & target partition\\ \hline
		$DQN_{scratch}$ & A DQN trained on a target partition from scratch\\ \hline
		$scan\_time$ & A time required to scan a dataset\\ \hline
		$\mathcal{X}$ & A set of extracted itemsets matching a target type\\ \hline
		$Z$ & A normalisation factor to compute $s_{k,m}$\\ \hline
	\end{tabular}
\end{table}

\bibliographystyle{unsrt}
\bibliography{refs}


\begin{IEEEbiography}[{\includegraphics[width=1in,height=1.25in,clip,keepaspectratio]{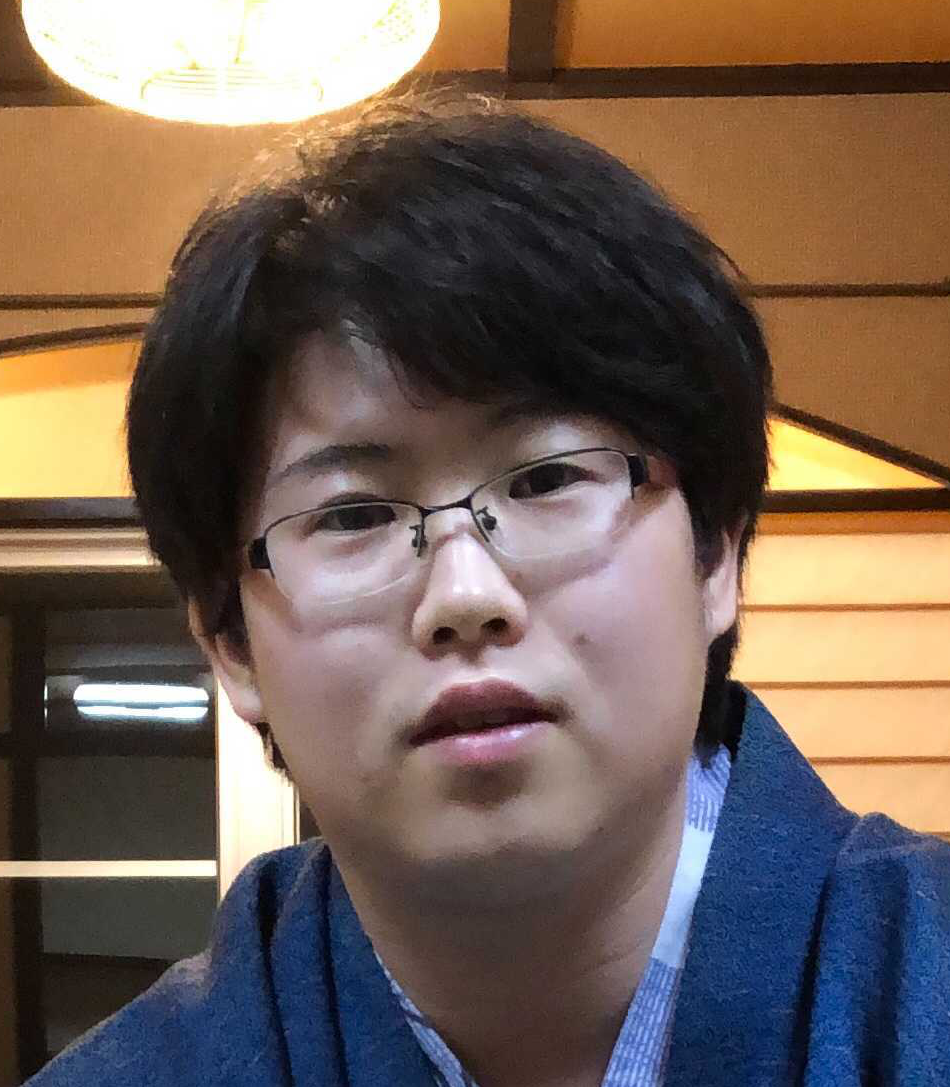}}]{Kazuma Fujioka} received his B.E. degree in Engineering from Kindai University, Japan in 2021. He is now pursuing his M.E. degree at Graduate School of Science and Engineering, Kindai University. His research interests include deep reinforcement learning and data mining.
\end{IEEEbiography}

\begin{IEEEbiography}[{\includegraphics[width=1in,height=1.25in,clip,keepaspectratio]{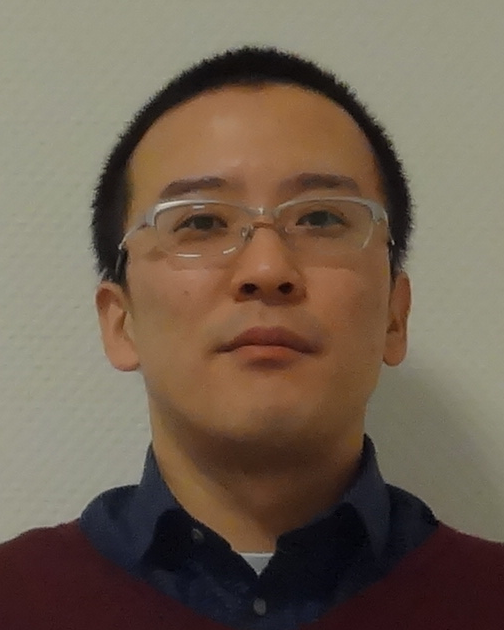}}]{Kimiaki Shirahama} received his B.E., M.E. and D.E degrees in Engineering from Kobe University, Japan in 2003, 2005 and 2011, respectively. After working as an assistant professor in Muroran Institute of Technology, Japan, he worked as a postdoctoral researcher at Pattern Recognition Group in University of Siegen, Germany from 2013 to 2018. Since 2018, he is working as an associate professor at Kindai University, Japan. His research interests include multimedia data processing, machine learning, data mining and sensor-based human activity recognition. He is a member of ACM SIGKDD, ACM SIGMM, the Institute of Image Information and Television Engineers in Japan (ITE), Information Processing Society of Japan (IPSJ) and the Institute of Electronics, Information and Communication Engineering in Japan (IEICE).
\end{IEEEbiography}

\EOD

\end{document}